\documentclass[fleqn,usenatbib]{mnras}

\usepackage{newtxtext,newtxmath}
\usepackage[T1]{fontenc}
\usepackage{gensymb}
\usepackage{graphicx}
\usepackage{subcaption}
\usepackage{caption}
\usepackage{tabularx}
\usepackage{booktabs}
\usepackage{makecell}
\usepackage{threeparttable}

\usepackage[switch]{lineno} % 'switch' handles two-column layout better
\usepackage{xcolor}

\DeclareRobustCommand{\VAN}[3]{#2}
\let\VANthebibliography\thebibliography
\def\thebibliography{\DeclareRobustCommand{\VAN}[3]{##3}\VANthebibliography}

\title[SN~2022xus]{SN~2022xus: bridging the gap between Type ~IIP and IIL supernovae}

\author[M. Dubey et al.]{Monalisa Dubey,$^{1,2}$\thanks{E-mail: monalisa@aries.res.in, monalisadubeyprl@gmail.com}
Kuntal Misra,$^{1}$\thanks{E-mail: kuntal@aries.res.in, kuntalmishra@gmail.com}
Naveen Dukiya,$^{1,2}$
Kumar Pranshu,$^{1,3}$
Raya Dastidar,$^{4}$
\newauthor
K. Azalee Bostroem,$^{5,6}$
Yize Dong,$^7$
Joseph R. Farah,$^{8}$
Estefania Padilla Gonzalez,$^{9}$
Emily Hoang,$^{10}$
\newauthor
D. Andrew Howell,$^{8,11}$
Curtis McCully,$^{8,11}$
Megan Newsome,$^{12}$
Craig Pellegrino, $^{13}$
Aravind Pazhayath Ravi,$^{10}$
\newauthor
Nicol\'as Meza Retamal,$^{10}$
Ajay Kumar Singh,$^{14}$,
Giacomo Terreran$^{15}$ and
Stefano Valenti$^{10}$
\\
$^1$Aryabhatta Research Institute of Observational Sciences, Nainital-263001, India\\
$^2$Mahatma Jyotiba Phule Rohilkhand University, Bareilly-243006, India\\
$^3$Department of Applied Optics and Photonics, University of Calcutta, Kolkata, 700106, India\\
$^4$Istituto Nazionale di Astrofisica, Osservatorio Astronomico di Brera, via E. Bianchi 46, 23807 Merate (LC), Italy\\
$^5$Steward Observatory, University of Arizona, 933 North Cherry Avenue, Tucson, AZ 85721-0065, USA\\
$^6$LSST-DA Catalyst Fellow\\
$^7$Center for Astrophysics | Harvard $\&$ Smithsonian, 60 Garden Street, Cambridge, MA 02138-1516, USA\\
$^8$Department of Physics, University of California, Santa Barbara, CA 93106-9530, USA\\
$^9$Johns Hopkins University, San Martin Dr, Baltimore, MD 21210, USA\\
$^{10}$Department of Physics and Astronomy, University of California, Davis, 1 Shields Avenue, Davis, CA 95616-5270, USA\\
$^{11}$Las Cumbres Observatory, 6740 Cortona Drive, Suite 102, Goleta, CA 93117-5575, USA\\
$^{12}$NASA Goddard Space Flight Center, 8800 Greenbelt Rd, Greenbelt, MD 20771, USA\\ 
$^{13}$Department of Astronomy, The University of Texas at Austin, 2515 Speedway, Stop C1400, Austin, TX 78712, USA\\
$^{14}$Department of Applied Physics/Physics, Bareilly College, Mahatma Jyotiba Phule Rohilkhand University, Bareilly, Uttar Pradesh-243001, India\\
$^{15}$Adler Planetarium, 1300 S DuSable Lake Shore Dr, Chicago, IL 60605, USA
}

\begin{document}
\label{firstpage}
\pagerange{\pageref{firstpage}--\pageref{lastpage}}
\maketitle

%\linenumbers
\begin{abstract}
We present optical photometric and spectroscopic observations of the Type~II supernova SN~2022xus. The SN reached its peak {\em V} band magnitude of  $-16.32$ mag within $\sim$7 days of explosion, followed by a plateau phase lasting $\sim$94 days with a declination rate of $\sim$1.2 mag (100 day)$^{-1}$. Early time spectra exhibit broad features that could be caused by the blending of several high-ionisation lines, likely arising from a relatively weak interaction between the SN ejecta and the surrounding circumstellar medium (CSM). Compared to typical Type~IIP SNe, SN~2022xus exhibits a smaller H$\alpha$ absorption-to-emission ratio ($a/e$), indicating a relatively small hydrogen envelope mass at the time of explosion. From nebular-phase spectroscopy and bolometric light curve modelling, the progenitor mass is estimated to be in the range of 12 -- 15 M$_\odot$. The multi-band light curve modelling using \texttt{REDBACK} infers a similar progenitor mass, a low mass-loss rate, and a confined CSM. Although several photometric and spectroscopic characteristics place the SN within the Type~IIL population, it displays mixed properties of both Type~IIP and Type~IIL SNe and cannot be cleanly classified into either subclass. We therefore identify SN~2022xus as a transitional event between Type~IIP and Type~IIL SNe, providing further evidence for a continuum between these two classes.

\end{abstract}


%%%%%%%%%%%%%%%%%%%%%%%%%%%%%%%%%%%%%%%%%%%%%%%%%%

%%%%%%%%%%%%%%%%% BODY OF PAPER %%%%%%%%%%%%%%%%%%
%\linenumbers
\section{Introduction:}
\label{sec:intro}
Supernovae (SNe) that exhibit strong Balmer lines in their spectra are classified as Type~II SNe (SNe~II henceforth), produced by the collapse of massive stars \citep[$>$ 8 M$_\odot$,][]{Filippenk_1997}. SNe~II are further divided into subclasses: SNe~IIP and IIL are classified based on light curve morphology, while SNe~IIn and IIb are distinguished based on their spectroscopic characteristics. SNe~IIP show nearly constant luminosity for a span of a few months, resulting in a `plateau' in their light curve, whereas SNe~IIL show a more steeply declining light curve \citep{Barbon_1979}. The progenitors of SNe~IIP have been identified to be Red Super Giant (RSG) stars of $\sim$8--17 M$_\odot$ from direct imaging \citep{smart_2009, smart_2015}; whereas there are no confirmed direct detections of the progenitors of SNe~IIL \citep{VanDyk2017}. SNe~IIL are thought to be occurring from more massive progenitors, which suffer a higher mass loss; hence, they exhibit a low hydrogen (H) mass before the explosion \citep{Gutierrez_2014, IIP_IIL_2021dbg}. 

Generally, SNe~IIL are distinguished based on their rapid decline in the light curve during the photospheric phase \citep[$>$ 0.5 mag (50 day)$^{-1}$,][]{Faran_2014}. \cite{Patat_1994} found that the mean apparent brightness of SNe~IIL is $\sim$1.5 mag brighter than SNe~IIP. Their H$\alpha$ absorption to emission ($a/e$) ratio is also shallower than that of SNe~IIP. Strong interaction between the SN ejecta and circumstellar material (CSM) has been invoked for several SNe~IIL (e.g., SN~2008es, \citealt{IIL_2008es}; ASASSN-15oz, \citealt{IIL_ASASSN-15oz}; SN~2016iog, \citealt{IIL_2016iog}), primarily on the basis of their high luminosities, which are difficult to explain without an additional radiation source. The narrow high-ionisation emission lines that would directly confirm CSM interaction were not detected, most likely because sufficiently early spectroscopy was not obtained. However, early spectra of several SNe~IIP (e.g., SNe~2021gmj; \citealt{Murai_2021gmj}, 2021yja; \citealt{griffin_2021yja}, 2023ixf; \citealt{IIP_2023ixf, Bostroem_2023ixf, Zhang_2023ixf, Teja_2023ixf, Smith_2023ixf, Zimmerman_2023ixf, Singh_2023ixf}, 2024ggi; \citealt{IIP_2024ggi, Zhang_2024ggi, Jacobson_2024ggi, Pessi_2024ggi}, 2024bch; \citealt{IIP_2024bch, Andrews_2024bch}) do reveal these signatures directly, providing unambiguous evidence for CSM--ejecta interaction at very early epochs. Some SNe exhibit blended characteristics of SNe~IIP and IIL and are interpreted as a bridge between the two classes (e.g., SNe~2013ej; \citealt{bose_2013ej, 2013ej_Mauerhan}, 2021dbg; \citealt{IIP_IIL_2021dbg}). Furthermore, there are some SNe identified as transitional candidates between SNe~IIP and IIn, which are thought to arise from high-mass stars, such as luminous blue variables (LBVs), and exhibit intense CSM interaction (e.g., PTF11iqb; \citealt{Smith_PTF11iqb}, SNe~2013fs; \citealt{Bullivant_2018}, 2020pvb; \citealt{Elias_2020pvb}). The continuity between SNe~IIP and IIL is well established by large statistical studies of SNe~II \citep{Anderson_2014, Sanders_2015, Valenti_2016, Gutierrez_2014, Gutierrez_2017}, which find no clear evidence that the two types of SNe arise from distinct progenitor populations. Brighter SNe~II exhibit higher photospheric velocity and produce more $^{56}$Ni \citep{Hamuy_2002, Hamuy_2003}, as well as the surrounding CSM density, which can significantly influence the observed characteristics of the SN \citep{Morozova_2017, Gutierrez_2014}. Therefore, the initial mass of the progenitor, mass-loss history, and nearby CSM may primarily influence the diversity in the SNe~II population.

In most studies, hydrostatic equilibrium is assumed to simulate H-rich SNe; however, because of an extended convective H-rich envelope, RSG stars exhibit brightness variations attributed to radial pulsations \citep{Bono_2000, Kiss_2006, Goldberg_2020}. The effect of stellar pulsation has been studied in nearby SNe (e.g., SNe~2023ixf; \citealt{Niu_2023, Kilpatrick_2023ixf, Jencson_2023ixf, Soraisam_2023ixf}, 2024ggi \citealt{Xiang_2024}). Pulsation can introduce diversity in H-rich SNe~II light curves and can also produce early excess emission without the need for CSM surrounding the SN \citep{Bronner_2025, Laplace_2025, Suzuki_2025}.

\begin{figure*}
    \centering
    \includegraphics[width=\linewidth]{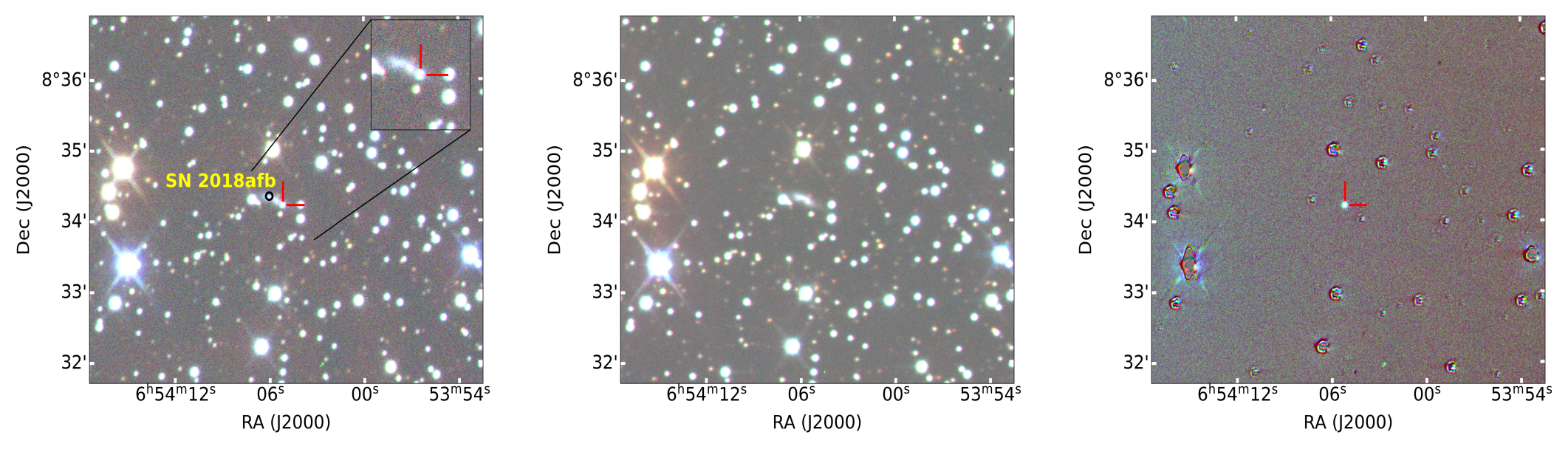}
    \caption{\textbf{Left Panel:} Colour composite image, created using SDSS {\em g}, {\em r}, and {\em i} filters, of SN~2022xus observed on UTC 2022-10-20.22. The SN is marked in red crosshairs, along with another SN~2018afb, previously discovered in the same host galaxy. The zoomed-in view of the SN is shown in the inset. \textbf{Middle Panel:} Reference image observed on UTC 2024-03-29.99. \textbf{Right Panel:} The host galaxy subtracted image of the SN marked with the red lines.}
    \label{fig:SN_location}
\end{figure*}

In this work, we study SN~2022xus (internal name: ATLAS22biqa), which exhibits characteristics of both SNe~IIP and IIL. The SN was discovered by the Asteroid Terrestrial-impact Last Alert System (ATLAS) group on UTC 2022-10-16.58 (MJD 59868.58) using {\em orange}-ATLAS filter mounted on the ATLAS Haleakala telescope \citep{2022xus_ATLAS} with a magnitude of 17.92 (ABmag), and classified as a SN~II \citep{classification_2022xus}. It exploded at the outskirts of the galaxy LEDA~136560. The redshift ($z$ = 0.00875$\pm$0.00001\footnote{\url{https://ned.ipac.caltech.edu/}}) of the SN corresponds to a distance of 37.43$\pm$2.63 Mpc (corrected for Virgo + GA + Shapley cluster), calculated assuming H$_0 = 73$ km/s/Mpc, $\Omega_{\mathrm{matter}}=0.27$, and $\Omega_{\mathrm{vaccum}}=0.73$. The location of the SN (RA = 06:54:05.14, DEC = 08:34:13.37) within the host galaxy is marked in \autoref{fig:SN_location}, along with one more SN~II (SN~2018afb), which exploded earlier in the same galaxy.

The last non-detection by ATLAS was reported on UTC 2022-10-14.63 (MJD 59866.63), with a limiting magnitude of 19.03 (AB mag). There is nearly two days difference between the last non-detection and SN discovery. Hence, the midpoint of these two epochs is taken as the explosion epoch, $t_0 = \mathrm{MJD~} 59867.61\pm0.97$, and the half of the difference between the non-detection and discovery epochs is considered as the corresponding uncertainty. The \ion{Na}{}{ID} line is not detected at the redshift of the host galaxy in the spectra; therefore, only line of sight Galactic reddening \citep[{\em E(B-V)$_{\rm MW}$}= 0.193 mag;][]{Schlafly_2011} and \cite{Cardelli_1989} extinction law is considered throughout the analysis. The basic information of SN~2022xus and its host galaxy is given in \autoref{tab:SN_info}.

The paper is structured as follows: the photometric and spectroscopic observations are detailed in \autoref{sec:Obs}. Spectroscopic analysis during different phases of the SN, along with spectral modelling, is discussed in \autoref{sec:spectra_analysis}. The light curve parameters and progenitor properties are presented in \autoref{sec:phot_analysis}. The placement of the SN within the diversity of SNe~II is discussed in \autoref{sec: diversity}. Finally, the findings of our study are summarised in \autoref{sec:summary}. 

\begin{table}
	\begin{center}
	   \caption{Basic information on SN~2022xus and LEDA~136560.}
	   \begin{tabular}{ll} 
		\hline
            \multicolumn{2}{c}{\textbf{SN~2022xus}} \\ \hline
        SN type & Type IIP\\
		RA & $06:54:05.139$\\
        Dec & $08:34:13.37$ \\
        Offset from nucleus &  8.5W 4.7S\\
	    Discovery date (MJD) & 59868.58\\
        Last non-detection (MJD) & 59866.63\\
        Estimated explosion epoch (MJD) & 59867.61$\pm$0.97\\
        Distance (Mpc) & 37.43$\pm$2.63 \\
        Total extinction E(B-V) (mag) & 0.193\\
	   \hline\multicolumn{2}{c}{\textbf{LEDA 136560}$^\dagger$} \\ \hline
        Semi-major axis diameter (arcsec) & 15 \\
        Semi-minor axis diameter (arcsec) & 15\\
		Redshift & 0.00875$\pm$ 0.00001	\\
		Helio. Velocity (km s$^{-1}$) & 2625$\pm$4 \\
		\hline
	   \end{tabular}
	   \label{tab:SN_info}
    \end{center}
    $^\dagger${Taken from NASA/IPAC Extragalactic Database (NED).}\\ 
\end{table}

\section{Observations:}
\label{sec:Obs}
The photometric follow-up observations were initiated two days after the discovery by the Las Cumbres Observatory (LCO) telescope network \citep{Brown_2013} under the Global Supernova Project (GSP). The multi-band observations in the {\em UBgVri} filters were conducted until 177.51 days after the explosion. Image pre-processing (e.g., bias and dark removal, flat-field correction, astrometry) was performed using the \texttt{BANZAI} pipeline \citep{McCully_2018}. The host galaxy template images, taken on UTC 2024-03-29.99 by the 1-m LCO telescope, were subtracted from the science images using the image subtraction software \texttt{HOTPANTS} \citep{hotpants_becker_2015}, integrated into the LCO photometric pipeline. The science, template, and subtracted images are presented in \autoref{fig:SN_location}. Point spread function (PSF) photometry was done on the host-subtracted images using the PyRAF-based photometric pipeline \texttt{lcogtsnpipe}\footnote{\url{https://github.com/LCOGT/lcogtsnpipe/}} \citep{Valenti_2016}. The SN magnitudes obtained from LCO data are given in \autoref{tab:optical_phot}. 

SN~2022xus was also observed in the Johnson–Cousins {\em UBVRI} filters with the 1.3-m Devasthal Fast Optical Telescope (DFOT), located at Devasthal, ARIES, India \citep{Joshi_2022}. The host galaxy template images acquired on UTC 2025-11-18.97 were subtracted from the science frames using the custom-built \texttt{ILMTDiff} image-subtraction algorithm \citep{2025MNRAS.538..133P}. The photometry was performed on the host-subtracted images using a custom \texttt{Python} photometry pipeline (Dukiya et al., in preparation). The PSFs were derived using PSFEx \citep{psfex_ascl} from the science images, and the instrumental magnitudes were derived by fitting the detected sources with the PSF model using \texttt{photutils} \citep{larry_bradley_2024_photutils}. The {\em UBVRI} instrumental magnitudes were calibrated against the {\it Gaia} Synthetic Photometry Catalog \citep{gaia_dr3}. The SN magnitudes obtained from DFOT data are presented in \autoref{tab:DFOT}.

\begin{table*}
 \begin{center}
 \caption{Photometric magnitudes of SN~2022xus with the LCO telescopes.}
 \label{tab:optical_phot}
 \scalebox{1}{
 \begin{tabular}{@{}lcccccccc}
 \hline
 Date   &  MJD   &  Phase$^\dagger$  &  {\em U}  &  {\em B}  &  {\em V}  &  {\em g}  &  {\em r}  &  {\em i} \\
 (UT)  &  (days)  &  (days)  &  (mag) &  (mag)  &  (mag)  &  (mag)  &  (mag)  &  (mag)\\ 
 \hline
 2022-10-18.0.37 & 59870.37 & 2.76 & 17.13 ± 0.02 & 17.56 ± 0.02 & 17.41 ± 0.01 & 17.32 ± 0.01 & 17.36 ± 0.02 & -- \\
2022-10-20.0.22 & 59872.22 & 4.61 & 17.20 ± 0.03 & 17.54 ± 0.04 & 17.29 ± 0.02 & -- & 17.25 ± 0.03 & 17.1 ± 0.02 \\
2022-10-21.0.34 & 59873.34 & 5.73 & -- & 17.30 ± 0.03 & 17.14 ± 0.02 & 17.29 ± 0.07 & 17.08 ± 0.02 & 17.11 ± 0.01 \\
2022-10-21.0.49 & 59873.49 & 5.88 & -- & 17.30 ± 0.03 & 17.14 ± 0.02 & 17.29 ± 0.07 & 17.14 ± 0.01 & 17.11 ± 0.01 \\
2022-10-23.0.10 & 59875.10 & 7.49 & 16.92 ± 0.01 & 17.44 ± 0.02 & 17.14 ± 0.01 & 17.12 ± 0.03 & 17.09 ± 0.02 & 17.01 ± 0.02 \\
2022-10-30.0.39 & 59882.39 & 14.78 & 17.72 ± 0.03 & 17.72 ± 0.02 & 17.32 ± 0.02 & 17.42 ± 0.02 & 17.13 ± 0.04 & 17.06 ± 0.02 \\
2022-11-02.0.47 & 59885.47 & 17.86 & 18.24 ± 0.01 & 17.96 ± 0.04 & 17.34 ± 0.02 & 17.52 ± 0.02 & 17.12 ± 0.02 & 17.05 ± 0.02 \\
2022-11-06.0.66 & 59889.66 & 22.05 & 18.51 ± 0.25 & 18.15 ± 0.03 & 17.35 ± 0.03 & 17.65 ± 0.13 & 17.07 ± 0.05 & 17.24 ± 0.03 \\
2022-11-10.0.23 & 59893.23 & 25.62 & 19.01 ± 0.11 & 18.24 ± 0.06 & 17.45 ± 0.06 & 17.82 ± 0.04 & 17.17 ± 0.05 & 17.20 ± 0.01 \\
2022-11-14.0.24 & 59897.24 & 29.63 & 19.03 ± 0.11 & 18.55 ± 0.03 & 17.57 ± 0.03 & -- & 17.27 ± 0.04 & -- \\
2022-11-22.0.34 & 59905.34 & 37.73 & 19.84 ± 0.02 & 18.61 ± 0.03 & 17.69 ± 0.01 & 18.11 ± 0.02 & 17.41 ± 0.01 & 17.17 ± 0.01 \\
2022-12-01.0.35 & 59914.35 & 46.74 & 20.07 ± 0.13 & 18.75 ± 0.03 & 17.71 ± 0.02 & 18.37 ± 0.02 & 17.33 ± 0.02 & 17.29 ± 0.02 \\
2022-12-08.0.58 & 59921.58 & 53.97 & -- & 19.16 ± 0.13 & 17.80 ± 0.04 & 18.35 ± 0.05 & 17.44 ± 0.03 & 17.29 ± 0.04 \\
2022-12-20.0.31 & 59933.31 & 65.70 & 20.55 ± 0.16 & 19.18 ± 0.02 & 17.90 ± 0.01 & 18.43 ± 0.02 & 17.50 ± 0.02 & 17.33 ± 0.03 \\
2022-12-26.0.24 & 59939.24 & 71.63 & 20.81 ± 0.17 & 19.22 ± 0.02 & 17.95 ± 0.01 & 18.50 ± 0.02 & 17.54 ± 0.01 & 17.46 ± 0.02 \\
2022-12-31.0.88 & 59944.88 & 77.27 & -- & 19.80 ± 0.17 & 17.97 ± 0.03 & 18.81 ± 0.06 & 17.64 ± 0.02 & 17.57 ± 0.04 \\
2023-01-13.0.56 & 59957.56 & 89.95 & -- & -- & 18.32 ± 0.11 & -- & 18.46 ± 0.09 & -- \\
2023-01-20.0.20 & 59964.20 & 96.59 & -- & 20.92 ± 0.08 & 19.69 ± 0.04 & 20.32 ± 0.04 & 18.86 ± 0.02 & 18.76 ± 0.03 \\
2023-01-27.0.25 & 59971.25 & 103.64 & -- & 21.69 ± 0.11 & 20.32 ± 0.07 & 20.89 ± 0.06 & 19.52 ± 0.03 & 19.48 ± 0.04 \\
2023-02-06.0.56 & 59981.56 & 113.95 & -- & -- & 20.57 ± 0.12 & 21.15 ± 0.22 & 19.51 ± 0.04 & 19.33 ± 0.07 \\
2023-02-10.0.92 & 59985.92 & 118.31 & -- & 21.57 ± 0.20 & 20.51 ± 0.16 & 21.17 ± 0.16 & 19.63 ± 0.04 & 19.52 ± 0.09 \\
2023-02-14.0.89 & 59989.89 & 122.28 & -- & 22.01 ± 0.13 & 20.51 ± 0.07 & 21.11 ± 0.08 & 19.71 ± 0.04 & 19.64 ± 0.09 \\
2023-02-20.0.52 & 59995.52 & 127.91 & -- & 21.87 ± 0.14 & 20.41 ± 0.04 & 21.25 ± 0.08 & 19.69 ± 0.03 & 19.53 ± 0.05 \\
2023-02-25.0.12 & 60000.12 & 132.51 & -- & 21.50 ± 0.09 & 20.74 ± 0.09 & 21.41 ± 0.10 & 19.78 ± 0.04 & 19.79 ± 0.07 \\
2023-03-03.0.09 & 60006.09 & 138.48 & -- & -- & -- & -- & 20.01 ± 0.03 & -- \\
2023-03-18.0.81 & 60021.81 & 154.20 & -- & 21.94 ± 0.13 & -- & 21.18 ± 0.09 & 19.88 ± 0.04 & 19.96 ± 0.08 \\
2023-04-11.0.12 & 60045.12 & 177.51 & -- & 22.34 ± 0.14 & 21.16 ± 0.08 & 21.74 ± 0.13 & 20.15 ± 0.05 & 20.15 ± 0.11 \\
 \hline
 \end{tabular}
 }
 
 \end{center}

 $^\dagger$Phase relative to the explosion epoch (MJD 59867.61).
 \end{table*}
\begin{table}
 \begin{center}
 \caption{Photometric magnitudes of SN~2022xus with DFOT.}
 \label{tab:DFOT}
 \scalebox{0.85}{
 \begin{tabular}{@{}cccccc}
 \hline
 Phase$^\dagger$  &  {\em U}  &  {\em B}  &  {\em V}  &  {\em R}  &  {\em I} \\
 (days)  &  (mag) &  (mag)  &  (mag)  &  (mag)  &  (mag) \\
 \hline
 16.39 & 17.69$\pm$0.02 & 17.85$\pm$0.01 & 17.27$\pm$0.01 & 16.90$\pm$0.01 & 16.61$\pm$0.01\\

 28.39 & -- & 18.29$\pm$0.03 & 17.44$\pm$0.02 & 16.91$\pm$ 0.01 & 16.62$\pm$0.01\\

 31.39 & -- & 18.45$\pm$0.02 & 17.61$\pm$0.01 & 17.05$\pm$0.01 & 16.71$\pm$0.01\\

 42.39 & -- & 18.65$\pm$0.01 & 17.67$\pm$0.01 & 17.11$\pm$0.01 & 16.68$\pm$0.01\\

 49.39 & -- & 18.78$\pm$0.03 & 17.76$\pm$0.01 & 17.22$\pm$0.01 & -- \\

 76.39 & -- & -- & 18.03$\pm$0.01 & 17.49$\pm$0.01 & 16.96$\pm$0.01\\

 89.39 & -- & -- & 18.69$\pm$0.01 & 18.12$\pm$0.01 & 17.54$\pm$0.01\\

 93.39 & -- & -- & 19.13$\pm$0.02 & 18.63$\pm$0.02 & 17.93$\pm$0.02\\
 \hline

 \hline
 \end{tabular}
 }
 
 \end{center}

 $^\dagger$Phase relative to the explosion epoch (MJD 59867.61).
 \end{table}

The first spectrum was obtained 1.98 days after the explosion with the Supernova Integrated Field Spectrograph \citep[SNIFS;][]{Aldering_2007} mounted on the University of Hawaii 88" (UH88) telescope of Mauna Kea Observatory. The data was reduced and calibrated using the procedure described in \cite{Tucker_2022}. The poor dichroic correction region ($\sim$4920--5400 \AA) has been removed from the spectrum. Further follow-up spectroscopic observations were carried out with the FLOYDS spectrograph \citep{FLOYDS_2011} mounted on the 2-m LCO telescopes spanning from 3.44 to 99.41 days since explosion. Spectral reduction along with wavelength and flux calibrations were done using the \texttt{floydsspec}\footnote{\url{https://github.com/svalenti/FLOYDS\textunderscore pipeline/}} pipeline \citep{Valenti_2014}. Two spectra at 13.85 and 69.80 days were taken with the Kitt Peak Ohio State Multi-Object Spectrograph \citep[KOSMOS,][]{kosmos, kosmos_2} mounted on the ARC 3.5-m telescope \citep{APO_telescope} at the Apache Point Observatory (APO) in New Mexico. Data preprocessing and reduction were performed in a standard manner using \texttt{IRAF} \citep{Tody1986, Tody1993}. One late-time (422.39 day) nebular spectrum was acquired with the Low Resolution Imaging Spectrometer \citep[LRIS;][]{Oke_1995}, mounted on the 10-m Keck-I telescope. The spectrum was reduced using the \texttt{LPipe} software package \citep{Perley_LRIS}.

All spectra were photometrically scaled to {\em UBgVri} bands using \texttt{lightcurve-fitting}\footnote{\url{https://github.com/griffin-h/lightcurve_fitting}} \texttt{Python} package \citep{Griffin_2022}, and corrected for the redshift of the host galaxy. The log of spectroscopic observations is listed in \autoref{tab:spec_log}.

\begin{table}
 \begin{center}
 \caption{Log of Spectroscopic Observations.}
 \label{tab:spec_log}
 \scalebox{1}{
 \begin{tabular}{@{}cccc}
 \hline
 Phase$^\dagger$ & Telescope$+$Instrument & Resolution & Wavelength Range  \\
 (days) & & ($\lambda/\Delta\lambda$) & (\AA) \\
 \hline
 1.98 & UH88$+$SNIFS & 1000--1300 & 3400--10000\\
 3.44 & FTN+FLOYDS & 400--700 & 3300--10180\\
 9.50 & FTN+FLOYDS & 400--700 & 3300--10180\\
 13.85 & ARC$+$KOSMOS & $\sim$2100 & 3500--10000\\
 26.58 & FTS+FLOYDS & 400--700 & 3300--10180\\
 32.53 & FTS+FLOYDS & 400--700 & 3300--10180\\
 38.52 & FTS+FLOYDS & 400--700 & 3300--10180\\
 44.48 & FTS+FLOYDS & 400--700 & 3300--10180\\
 52.47 & FTS+FLOYDS & 400--700 & 3300--10180\\
 58.51 & FTS+FLOYDS & 400--700 & 3300--10180\\
 64.52 & FTN+FLOYDS & 400--700 & 3300--10180\\
 69.80 & ARC$+$KOSMOS & $\sim$2100 & 3500--10000\\
 70.52 & FTS+FLOYDS & 400--700 & 3300--10180\\
 80.47 & FTS+FLOYDS & 400--700 & 3300--10180\\
 99.41 & FTS+FLOYDS & 400--700 & 3300--10180\\
 422.39 & Keck$+$LRIS & 1200--2200 & 3200--10200\\

\hline
 \end{tabular}}
 
 \end{center}
$^\dagger$Phase relative to the explosion epoch (MJD 59867.61). 
\end{table}

\section{Spectral Analysis}
\label{sec:spectra_analysis}
\subsection{Early time CSM interaction signature}
\label{sec:early_spectra}

\begin{figure}
    \centering
    \begin{subfigure}[b]{\columnwidth}
        \centering
        \includegraphics[width=\columnwidth]{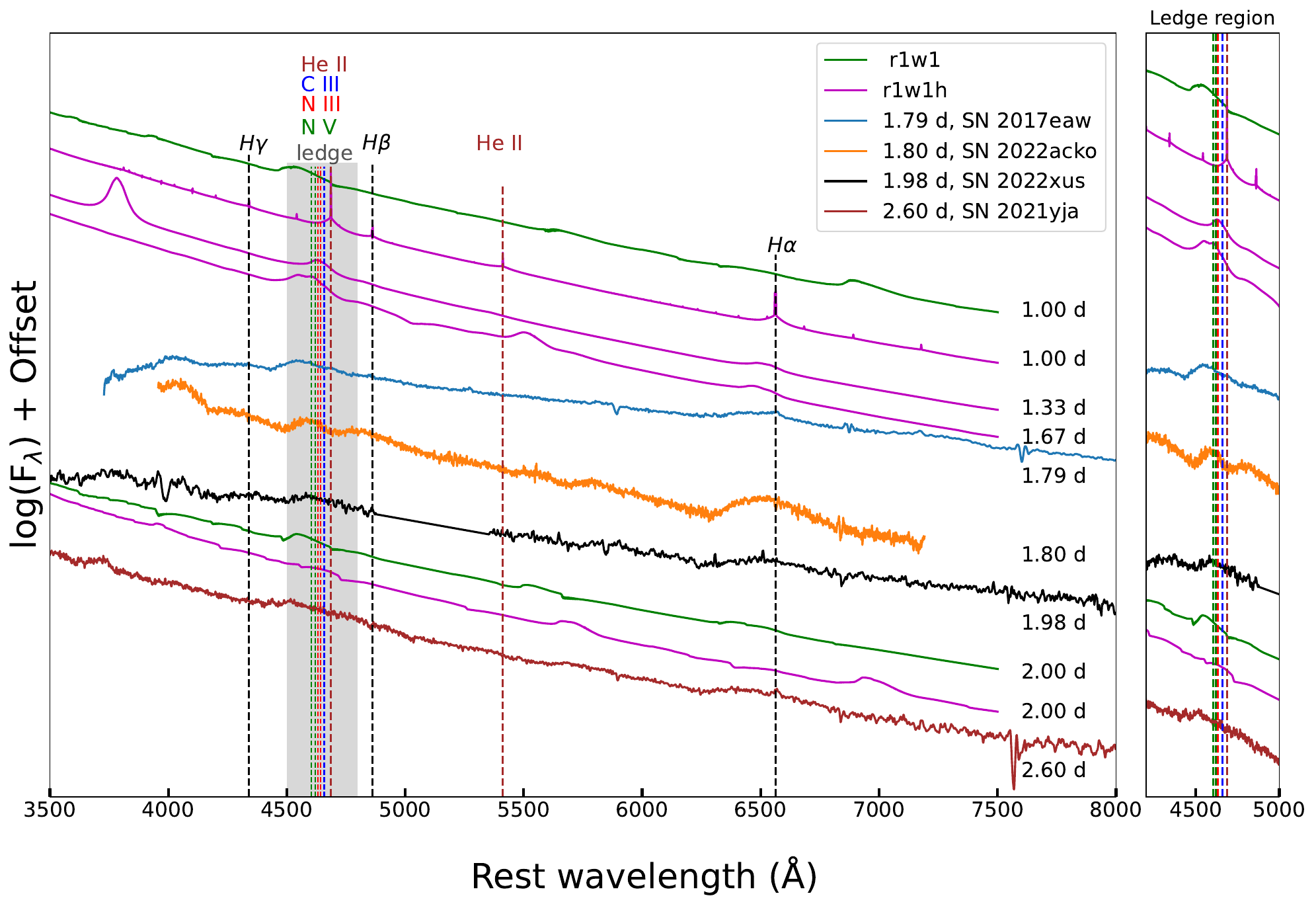}
    \end{subfigure}
    \hfill
    \begin{subfigure}[b]{\columnwidth}
        \centering
        \includegraphics[width=\columnwidth]{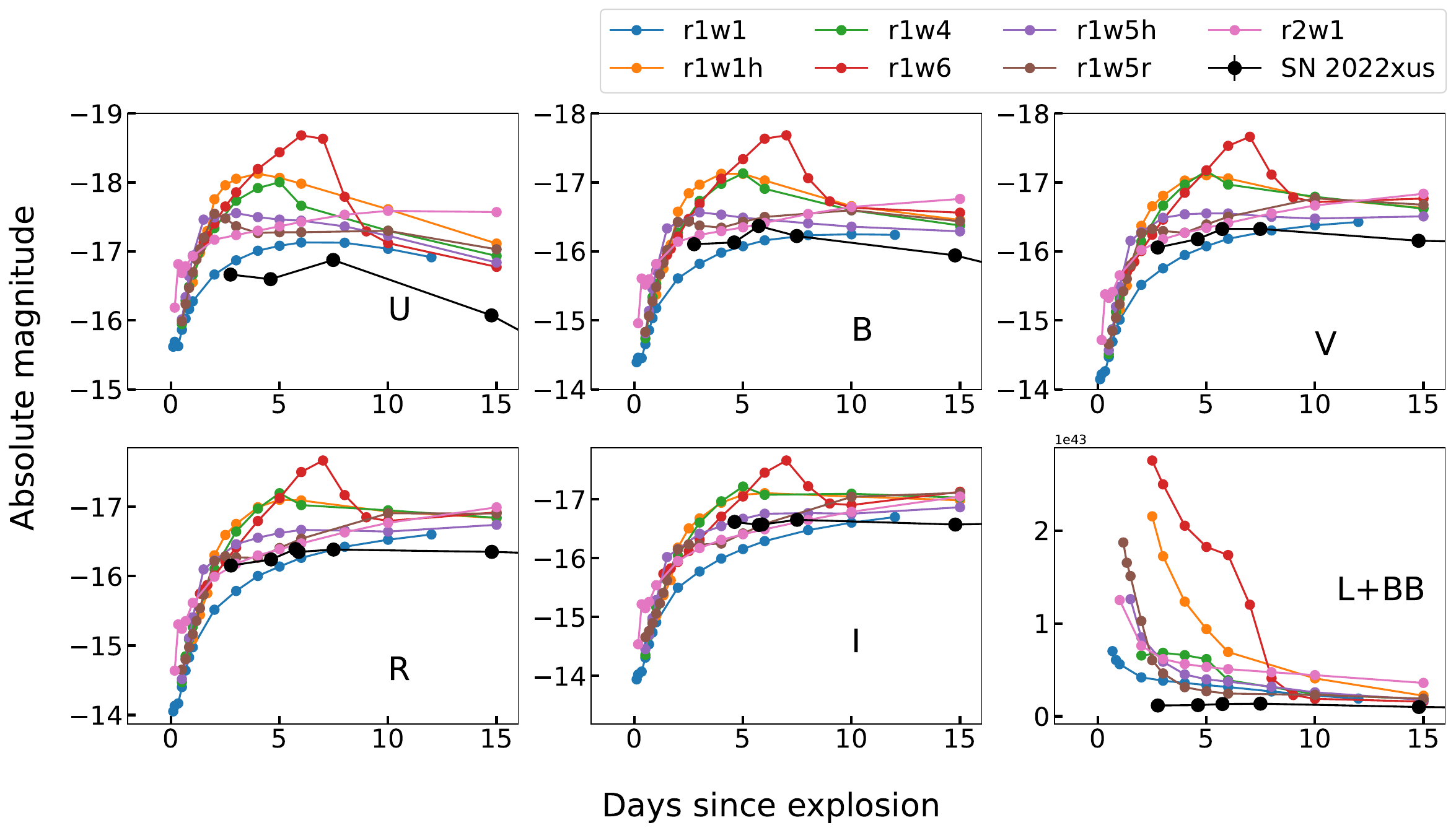}
    \end{subfigure}
    \caption{\textbf{Top Panel:} The earliest (1.98 day) spectrum of SN~2022xus is presented and compared with SNe~2017eaw, 2021yja, and 2022acko as well as with the model spectra computed for models \texttt{r1w1} and \texttt{r1w1h} \citep{Dessart_2017} at similar epoch. The broad H$\alpha$ profile is observed in both the observed and model spectra. An asymmetric `ledge-shape' feature spanning 4500–4800 \AA\ is also evident, as highlighted in the zoomed-in panel on the right. \textbf{Bottom Panel:} The multi-band absolute magnitude and bolometric light curve of SN~2022xus are compared with those of the models presented in \citet{Dessart_2017}.}
    \label{fig:dessert}
\end{figure}

The earliest spectrum of SN~2022xus, shown in \autoref{fig:dessert}, was acquired 1.98 days after the explosion. The spectrum is mostly blue and featureless, with no prominent spectral lines; however, a broad H$\alpha$ emission line with a shallow absorption dip is visible. \cite{JG_2024} suggested that these type of broad spectral features observed within hours to days after explosion, arise from Doppler broadening, as low-density, optically thin CSM ($\rho < 10^{-15}$ g cm$^{-3}$ at $R \approx 10^{15}$–$10^{16}$ cm) can be rapidly swept up by the fast-moving ejecta. The SN spectrum also exhibits an asymmetric `ledge' feature around 4500--4800 \AA. Some studies identify it as a blue-shifted component of \ion{He}{ii} 4686 \AA\ line originating from the SN ejecta \citep{Bullivant_2013fs}, while others suggest it may arise from the blending of weak high-ionisation lines \citep{Soumagnac_2020, Bruch_2021, griffin_2021yja, Lin_2022acko}. If the interaction between the CSM and the SN ejecta is not sufficiently strong to produce prominent high-ionisation lines, a broad feature may emerge as a result of line blending \citep{Lin_2022acko, Shrestha_2023axu, Dubey_2026}. Alternatively, this feature may also result from asymmetries in the CSM \citep{Andrews_2017gmr}. As early spectra of SN~2022xus do not show any prominent high-ionisation lines, the `ledge' feature may arise from the blending of several weak high-ionisation lines, a result of weak CSM interaction.

The early spectrum of SN~2022xus is compared with that of SNe~2017eaw \citep{szalai_2017eaw}, 2021yja \citep{griffin_2021yja}, and 2022acko \citep{Lin_2022acko}, which exhibit similar spectral features indicative of low-level CSM interaction. The SN spectrum is also compared with model spectra (\autoref{fig:dessert}; top panel) computed from different progenitor models \citep{Dessart_2017}. These spectra are generated using a 1D non-local thermodynamic equilibrium (NLTE) radiative transfer code, CMFGEN \citep{Dessert_CMFGEN_2012}. The model spectra computed from two models, \texttt{r1w1} (radius of the star, $R_\star$ = 501 R$_\odot$), and \texttt{r1w1h} ($R_\star$ = 1107 R$_\odot$, atmospheric height of 0.3 $R_\star$) which are simulated for 15 M$_\odot$ RSG progenitor with solar metallicity embedded in low density CSM (10$^{-6}$ M$_\odot$ yr$^{-1}$). The broad H$\alpha$ profile and `ledge' feature are reproduced by all model spectra apart from the 1 day spectrum of the \texttt{r1w1h} model, where high-ionisation lines are distinguishable, indicating a higher level of early CSM interaction. The observed {\em UBVRI} absolute magnitudes and the bolometric light curve, generated with \texttt{SuperBol} \citep{Nicholl_2018} using {\em UBgVri} filters, are compared with those of the models in \autoref{fig:dessert} (bottom panel). The early light curves of SN~2022xus evolve similarly to the \texttt{r1w1} model. Hence, the model comparison again suggests that SN~2022xus exploded from a compact RSG progenitor that was embedded in a low-density CSM.

\subsection{Spectral Evolution}
\label{sec:spectra_evo}

\begin{figure*}
    \centering
    \includegraphics[width=\textwidth]{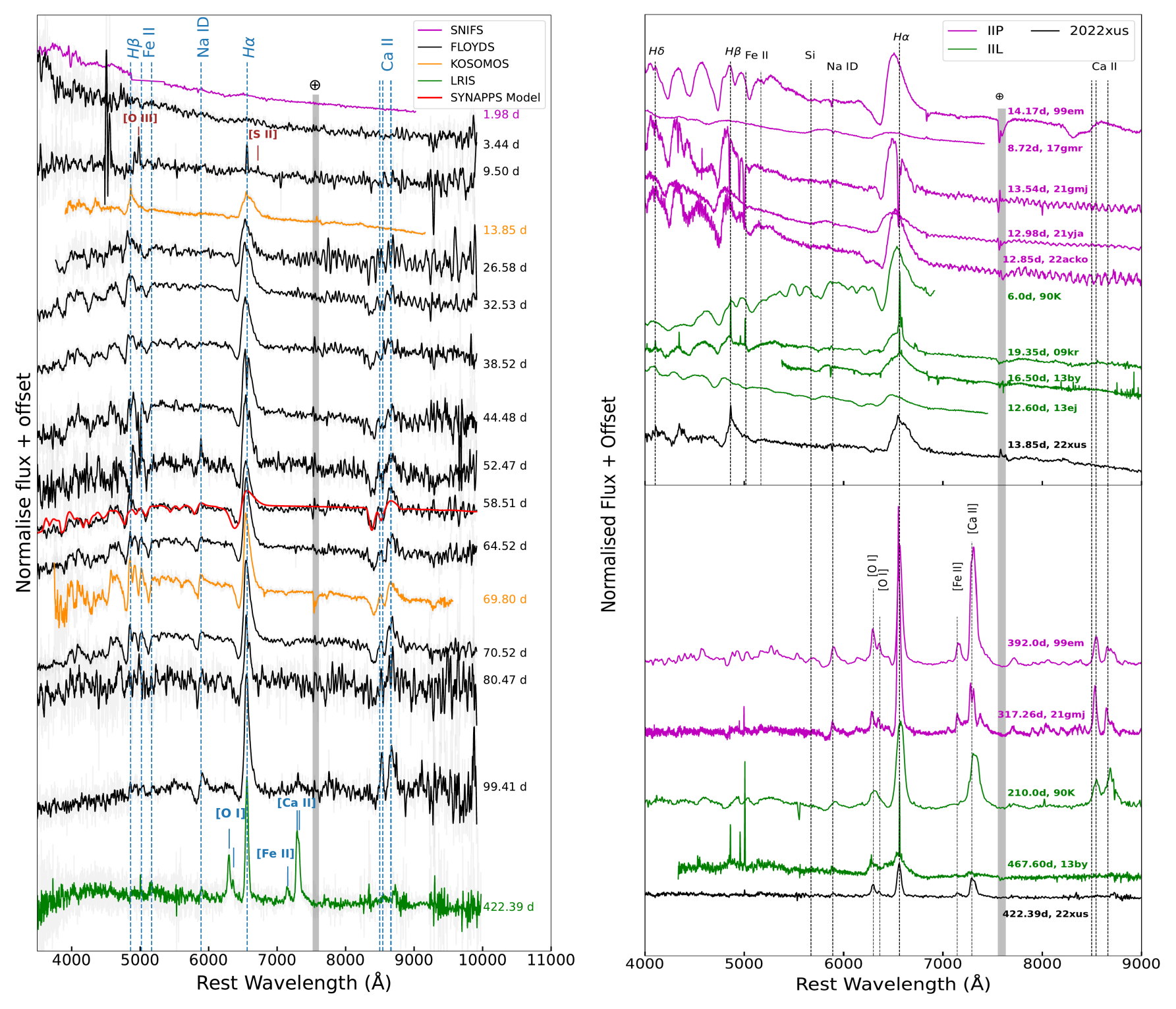}
    \caption{\textbf{Left Panel:} Spectroscopic evolution of SN~2022xus from 1.98 to 422.39 days since explosion. Host galaxy contamination is noticed in the 9.50 day spectrum. The model spectrum computed with SYNAPPS reproduced the 58.51 day spectrum shown in red. \textbf{Right Panel:} Comparison of the spectral feature of SN~2022xus with the comparison sample during the early-plateau (top), and nebular phase (bottom). The well-studied SNe~IIP and IIL samples are shown in magenta and green, respectively.}
    \label{fig:spec_ev}
\end{figure*}

The spectroscopic evolution of SN~2022xus spans from 1.98 to 422.39 days (in the observer frame of reference) and is presented in \autoref{fig:spec_ev} (left panel). The first two spectra are nearly featureless. In the 9.50 day spectrum, several galaxy lines (e.g., [\ion{O}{iii}] 3726, 3729 \AA, [\ion{S}{ii}] 6716, 6713 \AA) along with a narrow H$\alpha$ emission line appear due to contamination from the host galaxy. Broad Balmar line emissions are visible at 13.85 days. The P-Cygni profile of H$\alpha$ is prominent from 26.58 days onward. At the onset of the photospheric phase (32.53 day), metal lines (e.g., \ion{Fe}{ii} 5018, 5169 \AA, \ion{Ca}{ii} NIR triplet  8498, 8542, 8662 \AA, \ion{Na}{iD} 5893 \AA) start to become visible. In the optically thin and low-density nebular phase (422.39 day), the emission of H$\alpha$ is dominated by scattering. Several forbidden lines like [\ion{O}{i}] 6300, 6364 \AA\ doublets, [\ion{Fe}{ii}] 7155 \AA, and [\ion{Ca}{ii}] 7292, 7324 \AA\ are also visible. The strength of the \ion{Ca}{ii} NIR triplet line is very weak in this nebular spectrum.

To identify the elemental abundances in the SN ejecta during the photospheric phase, \texttt{SYNAPPS} \citep{Thomas_2011}, an open-source tool for forward modelling of SN spectra, is used to reproduce the 58.51 day spectrum. This model assumes spherically symmetric ejecta undergoing homologous expansion and a continuous blackbody spectrum emitted from the photosphere. Line formation is produced through resonance scattering and is treated under the Sobolev approximation \citep{Hummer_1985}. A set of parameters (e.g., different elements, effective temperature, photospheric velocity, opacity) within predefined ranges is supplied to the code as input to generate the synthetic spectra. By iteratively varying these parameters within their specified ranges, model spectra are generated that resemble the observed spectra. The model spectrum (in red) is overplotted on the 58.51 day spectrum in \autoref{fig:spec_ev} (left panel). The velocity of H$\alpha$ is estimated to be $\sim$ 5968 km s$^{-1}$ from the best-fit. The absorption depth of the H$\alpha$ profile is slightly overestimated, whereas H$\beta$ is well-matched by the model. This could be caused by the fact that both lines have different optical depths. The metal lines like \ion{Na}{I D}, \ion{Fe}{ii}, and \ion{Ca}{ii} are well-reproduced by the model. 

In \autoref{fig:spec_ev} (right panel), the early-plateau and nebular phase spectra of SN~2022xus are compared with a sample of SNe~IIP (SNe~1999em, 2017gmr, 2021gmj, 2021yja, and 2022acko) and SNe~IIL (SNe~1990K, 2009kr, 2013by, and 2013ej) (see \autoref{tab:comp_SN}). In the early-plateau phase, the SN exhibits broad H$\alpha$ emission with no absorption dip similar to other SNe~IIL such as SNe~2009kr, 2013by, and 2013ej. However, a similar profile is also observed for SNe~IIP such as SNe~2017gmr and 2021yja. SN~1990K exhibits a prominent absorption dip in the 6 day spectrum. During the nebular phase, the H$\alpha$ emission line, along with the forbidden lines, is visible, similar to other SNe; however, the Ca II NIR triplet at 8498, 8542, and 8662 \AA\ is not seen, like in SN~2013by. This may be attributed to the low density and temperature of the ejecta in the late nebular phase, which leads to a reduction in collisional excitation in atoms.

\begin{figure}
    \centering
    \includegraphics[width=\columnwidth]{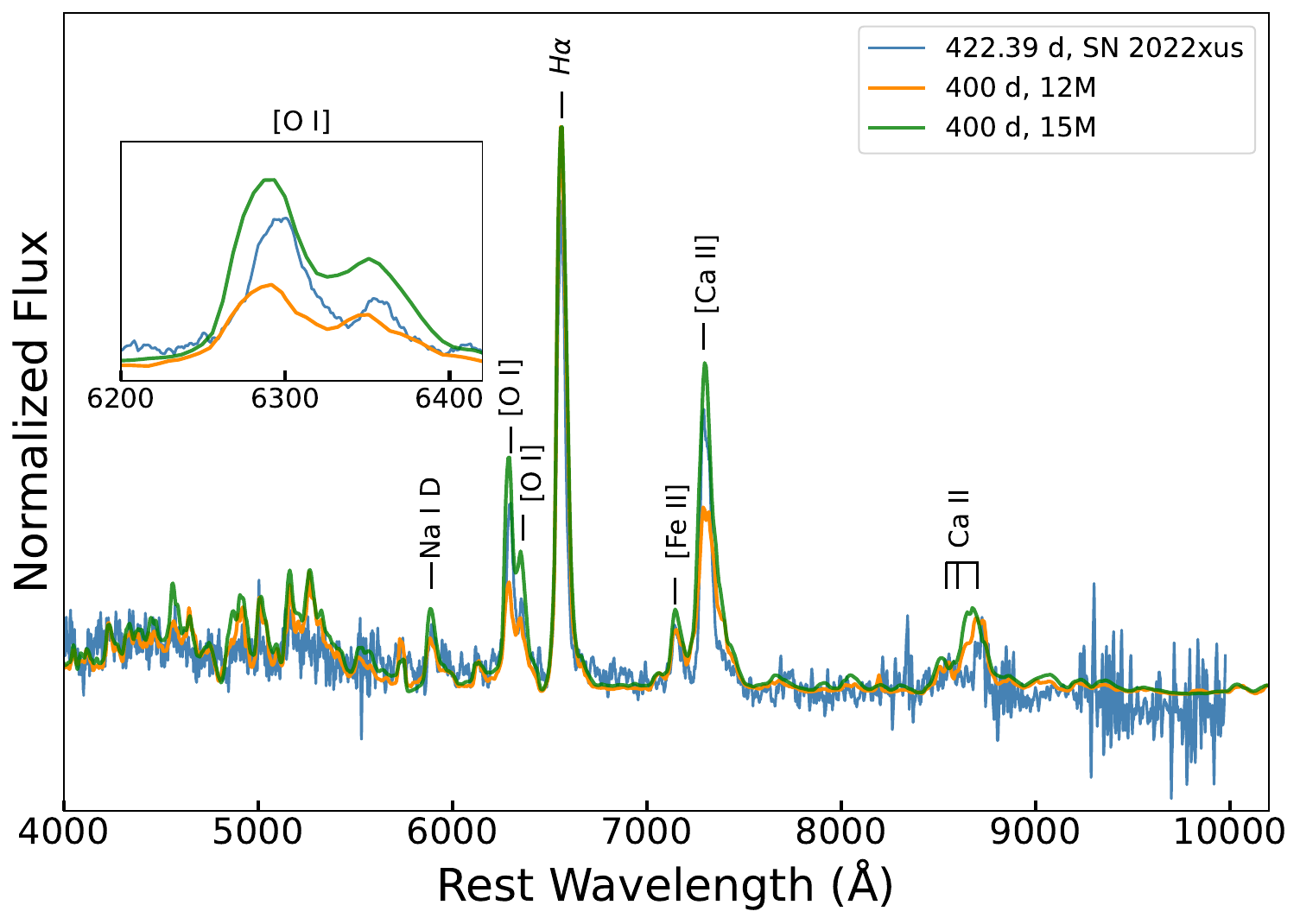}
    \includegraphics[width=\columnwidth]{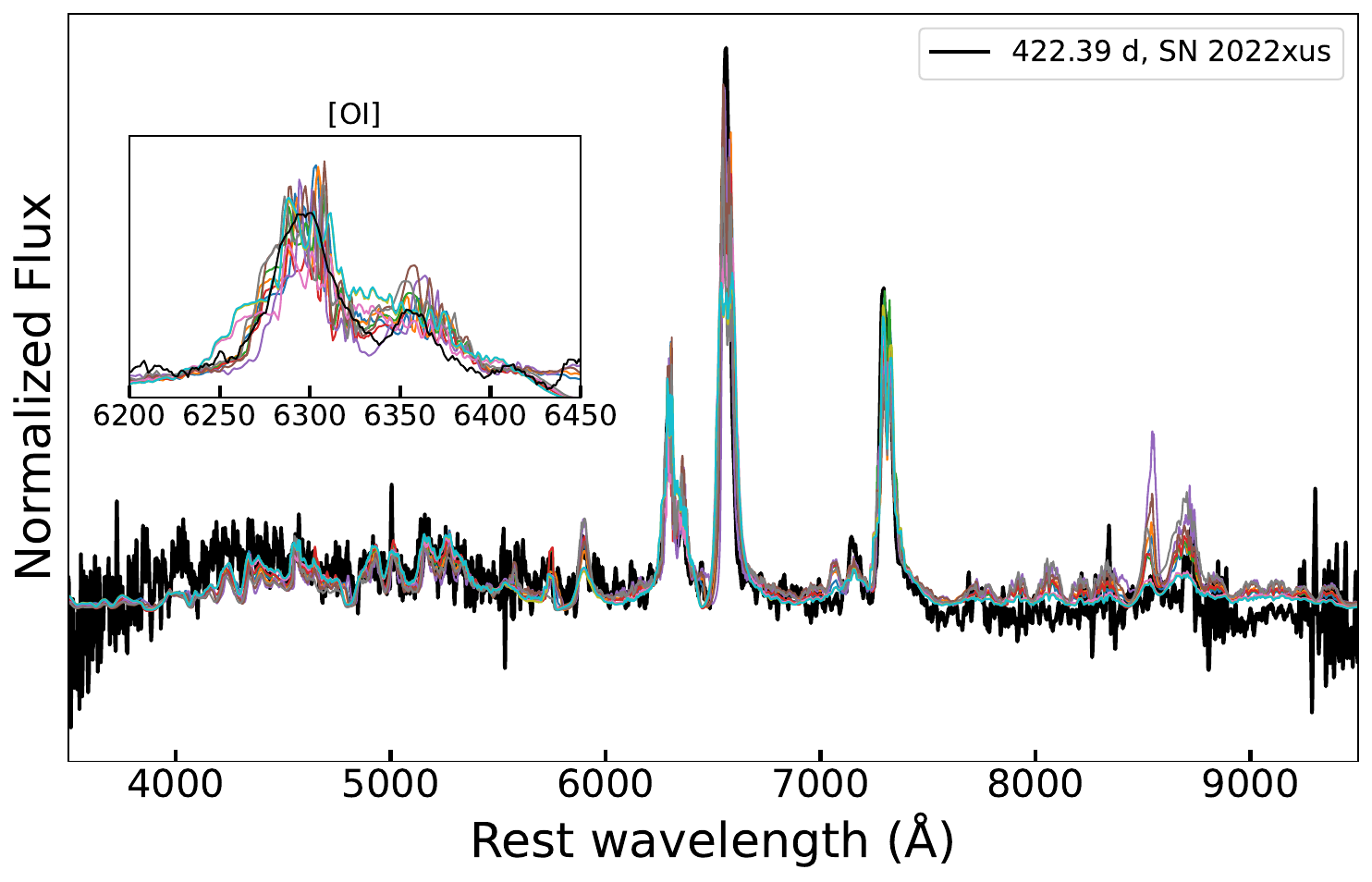}
    \caption{The nebular phase (422.39 day) spectrum of SN~2022xus is shown along with the model spectra computed for 12 and 15 M$_\odot$  \citep{Jerkstrand_2014} (top panel), and 10 best-fit model spectra (presented with different colours) from \citet{Dessart_2021} (bottom panel). The intensity of [\ion{O}{i}] 6300, 6364 \AA\ doublet of the observed and model spectra is presented in the inset.}
    \label{fig:neb_com}
\end{figure}

The zero-age main sequence (ZAMS) mass ($M_{ZAMS}$) of the progenitor star can be constrained using the [\ion{O}{i}] 6300, 6364 \AA\ doublets lines in the nebular phase \citep{Jerkstrand_2012, Jerkstrand_2014}. \cite{Jerkstrand_2014} computed synthetic nebular spectra for 12, 15, 19, and 25 M$_\odot$ progenitors at multiple epochs, using \texttt{KEPLER} code \citep{Woosley_2007}. The 422.39 day spectrum of SN~2022xus is compared with the two model spectra of 12 and 15 M$_\odot$ at 400 days (\autoref{fig:neb_com}, top panel). To match the integrated flux, the model spectra are rescaled to the observed spectrum as described in \cite{Bostroem_2019}. This comparison infers that the progenitor mass may likely fall between 12 and 15 M$_\odot$. 

To account for chemical mixing in the grid-based radiative transfer code, \citet{Dessart_2020} introduced a new technique to synthesise the nebular spectra of SNe~II acquired one year after explosion. In this 1D method, in a homologous expanding ejecta, the spherical shells containing distinct compositions are shuffled while preserving the original composition of each shell. In \citet{Dessart_2021}, they extended the previous study to a larger grid of progenitor mass (9--29 M$_\odot$), taking the models from \citet{Woosley_2007} (\texttt{WH07} henceforth) and \citet{Sukhbold_2016} (\texttt{S16} henceforth). The nebular spectra models\footnote{\url{https://zenodo.org/records/5525021}} computed in \citet{Dessart_2021} are compared to the observed nebular spectrum of SN~2022xus. Based on the root mean square error (RMSE) comparison of [\ion{O}{i}] 6300, 6364 \AA\ lines, the 10 best-fitted model spectra are plotted along with the observed spectrum in \autoref{fig:neb_com} (bottom panel). The comparison of [\ion{O}{i}] lines is given in the inset. Details of the 10 best-fitted model spectra are summarised in \autoref{tab:des_models}. From the best-fit models, the mass range is constrained to 12--15 M$_\odot$, which is consistent with the progenitor mass inferred from the nebular models of \citet{Jerkstrand_2014}. The energy range (0.34--0.61 $\times 10^{51}$ erg) obtained from the \texttt{S16} models, which is also in good agreement with the explosion energy estimated from the multi-band light curve modelling (see \autoref{sec:redback}). However, the $^{56}$Ni mass of the best-fit models spans 0.0317--0.0830 M$_\odot$, which is higher than the  $^{56}$Ni mass obtained in \autoref{sec:LC_evo}.

\begin{table}
\centering
\caption{The 10 best-fitting nebular models taken from \citet{Dessart_2021}.}
\label{tab:des_models}
\resizebox{\linewidth}{!}{%
\begin{tabular}{lcccc}
\toprule
Series $+$ Model & $M_{\rm ZAMS}$ ($M_\odot$) & $E_{\rm exp}$ ($10^{51}$ erg) & $M_{\rm Ni}$ ($M_\odot$) & RMSE \\
\midrule
S16 $+$ s12 (clumped) & 12.0  & 0.34  & 0.0317 & 0.0196 \\
S16 $+$ s12.5 (clumped) & 12.5 & 0.61  & 0.0507 & 0.0198 \\
WH07 $+$ s12A (ATOM) & 12.0 & 1.20 & 0.0504 & 0.0208 \\
WH07 $+$ s12A & 12.0 & 1.20 & 0.0504 & 0.0235 \\
S16 $+$ s12 & 12.0 & 0.34  & 0.0317 & 0.0243 \\
S16 $+$ s13.5 & 13.5 & 0.58  & 0.0593 & 0.0243 \\
WH07 $+$ s12A (ATOM) & 12.0 & 1.20 & 0.0504 & 0.0252 \\
WH07 $+$ s15A (ATOM) & 15.0 & 1.20 & 0.0830 & 0.0254 \\
S16 $+$ s12.5 (grid) & 12.5 & 0.61  & 0.0507 & 0.0255 \\
S16 $+$ s12.5 (grid+Fe\,I) & 12.5 & 0.61  & 0.0507 & 0.0262 \\
\bottomrule
\end{tabular}%
}
{\footnotesize ``ATOM'' indicates an updated atomic model, ``grid'' denotes a modified wavelength grid, and ``$grid+Fe\,I$'' includes an expanded $Fe\,I$ atomic dataset. The term ``clumped'' indicates a clumpy ejecta model.}
\end{table}

\begin{figure}
    \centering
    \includegraphics[width=\linewidth]{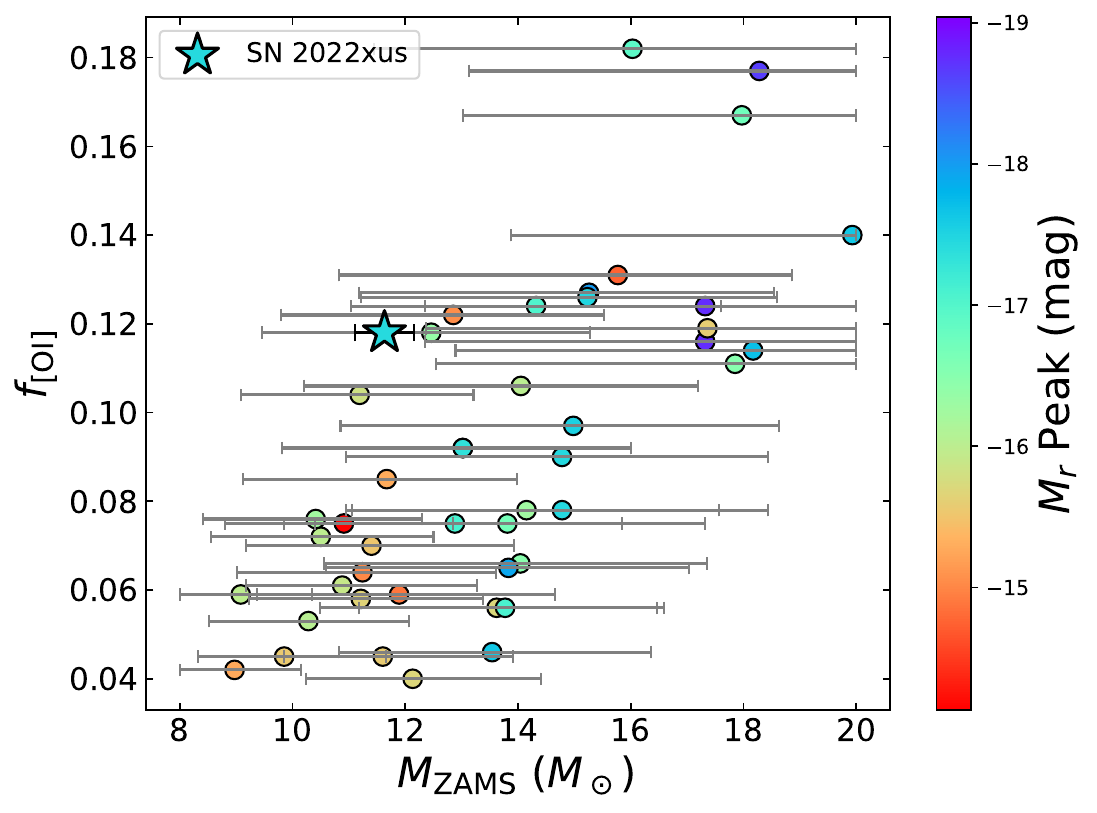}
    \caption{The comparison between $M_{ZAMS}$ and $f_{\rm [OI]}$ is presented for a sample given by \citet{Das_2026}. SN~2022xus is represented with a star marker. The corresponding M$_{r,peak}$ for each SNe represented by the colour bar.}
    \label{fig:fO_comp}
\end{figure}

The sample studies conducted on nebular spectra of SNe~II \citep{Jacobson_2025, Fang_2025, Das_2026} provide important information, such as the shape of H$\alpha$ profile during nebular phase can reveal sustained interactions between SN ejecta and CSM, constrain mass loss rate in pre-SN era and estimate progenitor mass as well as reveal the explosion mechanism from the observed strength of nebular lines (e.g, He- and O-shell emission line). The nebular spectrum of SN~2022xus exhibits a Gaussian H$\alpha$ profile, indicating minimal CSM interaction. Its progenitor mass (12--15 M$_\odot$), inferred from the nebular spectrum, is consistent with the low- to intermediate-luminosity RSG progenitor population identified for normal SNe~II. In \autoref{fig:fO_comp}, the $M_{ZAMS}$ and oxygen flux fraction ($f_{\rm [OI]}$) of the SN are plotted along with the sample taken from \citet{Das_2026}, who demonstrated that $f_{\rm [OI]}$ provides useful diagnostics of the oxygen-core mass and tried to identify the signature of electron capture SNe (ECSNe), which are expected to arise from ONeMg-core collapse of a $\approx$8--10 M$_\odot$ star. The corresponding peak magnitude in the {\em r} band for each SN is shown on the colour bar. Since SN~2022xus has only one nebular spectrum at 422.39 days, the variation of the nebular spectral line strength with phase could not be shown. For sample SNe, the $f_{\rm [OI]}$ values are adopted from nebular epochs closest to that of SN~2022xus. The $M_{ZAMS}$ of the SN is taken from the multi-band light curve modelling (see \autoref{sec:redback}). The increase of $f_{\rm [OI]}$ with $M_{ZAMS}$ implies that stars with higher initial masses produce larger oxygen cores, resulting in stronger [\ion{O}{i}] emission during the nebular phase. The nebular properties of SN~2022xus are also consistent with those of normal SNe~II, showing strong oxygen emission and more luminosity than expected for an ECSN.

\section{Light Curve Analysis}
\label{sec:phot_analysis}
\subsection{Shock-Cooling modelling}
\label{sec:shock_cooling}

\begin{figure}
    \centering
    \includegraphics[width=\columnwidth]{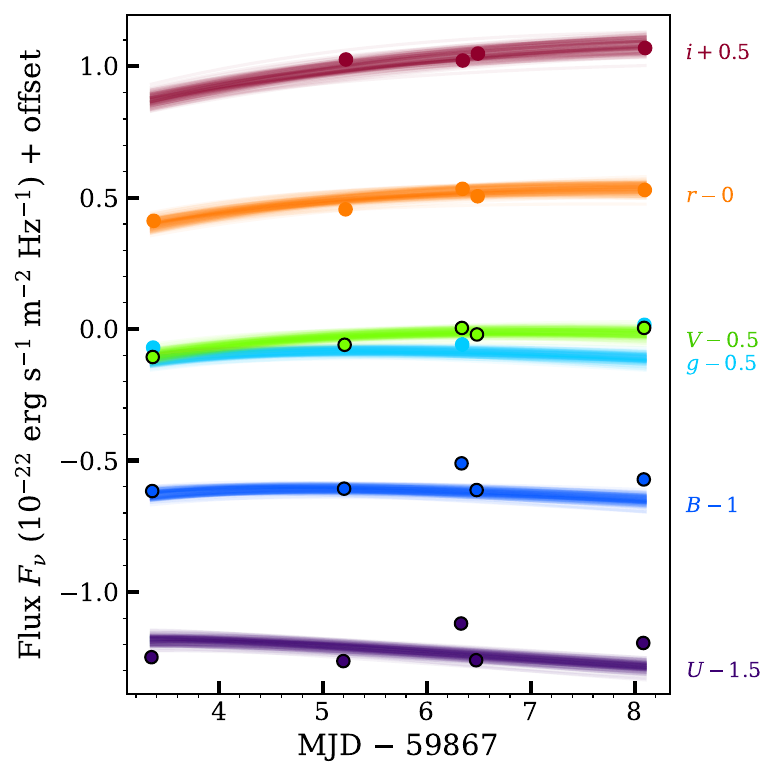}
    \caption{The early-phase light curve of SN~2022xus is modelled using shock-cooling method \citep{Sapir_Waxman_2017, hosseinzadeh_2020}. The observed light curves, along with the corresponding 100 best-fit models, are shown.}
    \label{fig:LC_model}
\end{figure}

In the absence of a dense CSM medium in the surroundings, the early light curves of SNe~II are generally powered by the shock-cooling mechanism. When the optical depth ahead of the shock, drops below $\approx c/v_s$ (where $v_s$ is the shock velocity), the shock breaks out at the stellar surface, producing an X-ray/UV flash (lasting seconds to minutes) followed by UV/optical emission produced from the cooling of the expanding envelope on a time scale of a day \citep{Waxman_2017}. The early {\em UBgVri} light curve ($<$ 10 days) of SN~2022xus is fitted using a shock-cooling model developed by \cite{Sapir_Waxman_2017}, implemented in the \texttt{Python}-based Light Curve Fitting routine by \cite{hosseinzadeh_2020}. This model estimates several physical parameters, including shock velocity ($v_s$), progenitor radius ($R$), envelope mass ($M_{env}$), distance ($d_L$), reddening (\textit{E(B-V)}), and explosion epoch ($t_0$). The fractional depth from the stellar surface is defined as $\delta \equiv \frac{R-r}{R}$, where $r$ is the radial coordinate. The $M_{env}$ is defined as the mass in the stellar envelope, where $\delta << 1$. Markov Chain Monte Carlo (MCMC) routine is employed for the fitting using a flat prior for all parameters except for $d_L$ and \textit{E(B-V)}, for which a Gaussian prior is used. The multiplication of ejecta mass and the numerical factor of the order unity ($f_\rho M$) is highly degenerate, and it cannot be constrained. The intrinsic scatter ($\sigma$) accounts for the additional photometric calibration and model uncertainties. The observed light curves, along with corresponding 100 best-fit models, are presented in \autoref{fig:LC_model}. The model shows good agreement with the observed light curves. The best-fit parameter values along with the priors used are provided in \autoref{tab:shock_cooling}. The model estimates $R\approx$ 230 R$_\odot$, which is within the typical radii of RSG stars \citep[100--1500 R$_\odot$,][]{Levesque_2017}; and it is also similar to the radius estimated by semi-analytical modelling (see \autoref{sec:LC_evo}). The estimated value of \textit{E(B-V)}, $d_L$, and t$_0$ also agree well with the values discussed in \autoref{sec:intro}. The model more strongly depends on $R$, and hence poorly constrains $v_s$ \citep{griffin_2021yja, Shrestha_2023axu, Hosseinzadeh_2023ixf} and, in this case, provides an unphysically lower value. Apart from this, the shock cooling modelling constrains all the estimated parameters well. 

\begin{table}
\centering
\begingroup
\footnotesize
\renewcommand{\arraystretch}{1.3} % Increase row spacing

\caption{Shock Cooling Parameters} 
\label{tab:shock_cooling}
\resizebox{\linewidth}{!}{%
\begin{tabular}{ccc}
\hline
\textbf{Parameter} & \textbf{Prior Range} & \textbf{Best-fit Value} \\
\hline
Shock Velocity [$v_{s}$ ($10^{8.5}$ cm/s)] & 0.0--3 & $0.69^{+0.07}_{-0.06}$  \\
Envelope Mass [$M_{\rm env}$ ( M$_\odot$)] & 0.0--10 & $6^{+3}_{-2}$  \\
Ejecta Mass $\times$ factor [$f_\rho M$ (M$_\odot$)] & 10.0--100.0 & $40\pm30$ \\
Progenitor Radius [$R$ ($10^{13}$ cm)] & 0.0--10.0 & $1.60^{+0.3}_{-0.2}$ \\
Distance [$d_L$ (Mpc)] & 30--50 & $35.04^{+0.06}_{-0.03}$ \\
Reddening [$E(B-V)$ (mag)] & 0.0--0.5 & $0.19^{+0.04}_{-0.03}$ \\
Explosion Time [$t_0$ (MJD)] & 59866--59870 & $59867.6^{+0.2}_{-0.3}$ \\
Intrinsic scatter [$\sigma$ ( Dimensionless)] & 0--10 & $7.9^{+1.1}_{-0.9}$ \\
\hline
\end{tabular}%
}
\endgroup
\end{table}

\subsection{Light curve evolution}
\label{sec:LC_evo}
\begin{figure}
    \centering
    \includegraphics[width=\columnwidth]{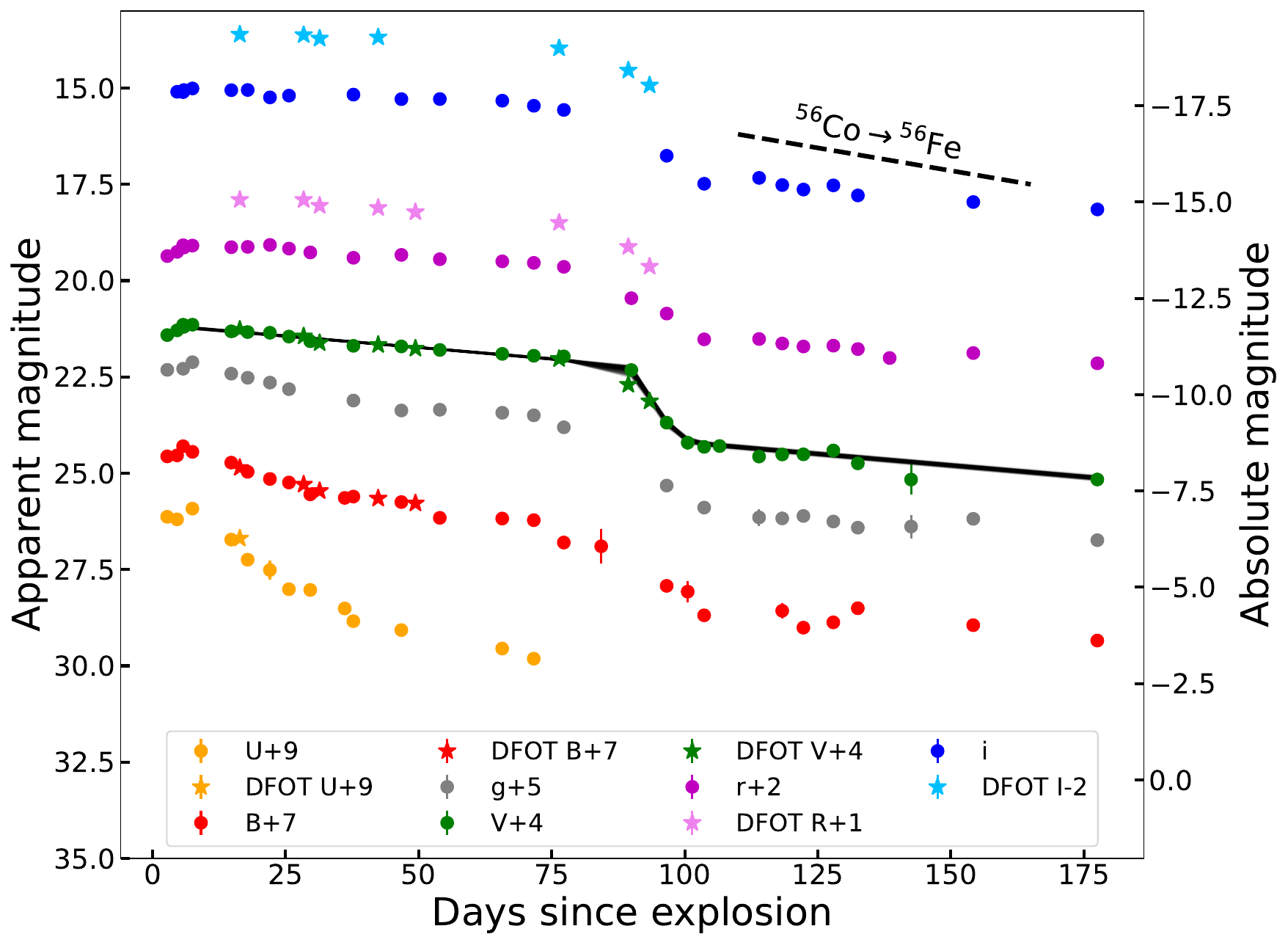}
    \caption{Optical light curves of SN~2022xus spanning from 2.76--177.51 days since explosion. The 100 best-fit models derived from the analytical modelling \citep{Valenti_2016} fitted to the {\em V} band light curve using MCMC, are shown in black.} 
    \label{fig:LC}
\end{figure}

The light curves of SN~2022xus in optical bands spanning from 2.76 to 177.51 days after explosion are presented in \autoref{fig:LC}. The peak absolute magnitude in the {\em V} band reaches $-16.32\pm0.01$ mag on 7.49 days after explosion. After the rise, the light curves remain approximately constant during the plateau phase with a decline rate ($S_{V}$) of 1.23$\pm$0.07 mag (100)$^{-1}$ days, then enter the nebular phase, marked by a drop in luminosity.

\begin{figure}
    \centering
    \includegraphics[width=0.95\columnwidth]{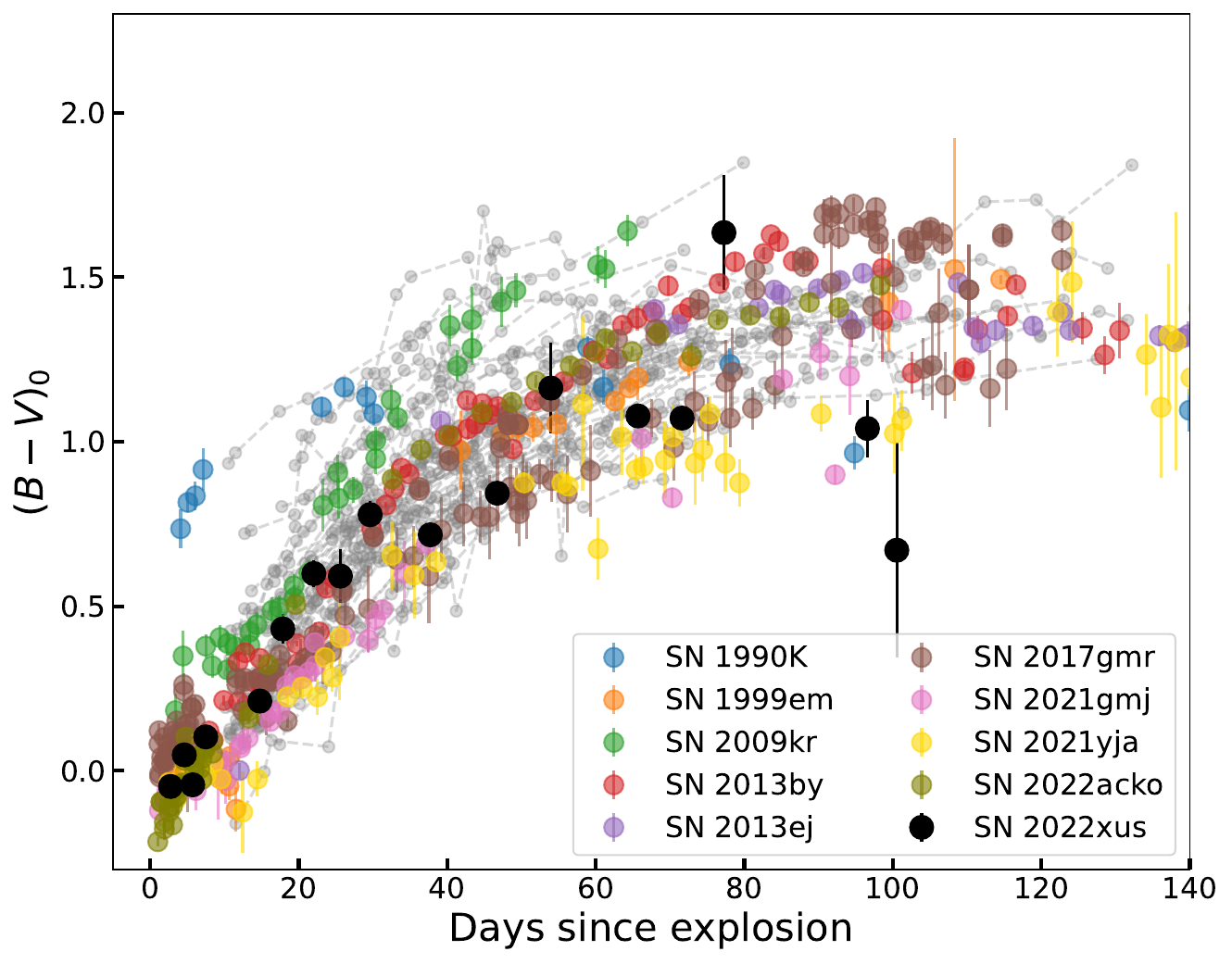}
    \caption{$(B-V)_0$ colour evolution of the SN is compared with the comparison sample SNe. Grey points represent the additional sample SNe taken from \citet{Jaeger_2018}.} 
    \label{fig:colour}
\end{figure}

To derive the observational parameters, the {\em V} band light curve is fitted using the analytical function given by \cite{Valenti_2016}. The fitting is performed with the MCMC sampler \texttt{emcee}\footnote{\url{https://emcee.readthedocs.io/en/stable/}}, which yields the following best-fit parameters: plateau length $t_{PT}$ = 94.80 $\pm$ 0.44 day, magnitude drop from plateau to nebular phase $a_0$ = 1.87 $\pm$ 0.03 mag, slope of the drop $w_0$ = 1.66 $\pm$ 0.38 mag/day, and value of slope before and after the drop $p_0 \approx 1.2$ mag (100 d)$^{-1}$. The 100 best-fit light curves to the {\em V} band (represented by black lines) are shown in \autoref{fig:LC}. \autoref{fig:colour} represents the $(B-V)_0$ colour evolution of the SN during the plateau phase, after which there is a scatter in colour. The SN colour is compared with a sample of well-observed events, which includes SNe~IIP such as SNe~1999em, 2017gmr, 2021gmj, 2021yja, and 2022acko, as well as SNe~IIL such as SNe~1990K, 2009kr, 2013by, and 2013ej, along with the additional sample taken from \citet{Jaeger_2018}. The light curve parameters of the comparison sample SNe are listed in \autoref{tab:comp_SN}.

The synthesised $^{56}$Ni mass can be estimated from the tail luminosity, as during the nebular phase, the energy input from the radioactive decay chain ($^{56}$Ni $\rightarrow$ $^{56}$Co $\rightarrow$ $^{56}$Fe) becomes the main driving force. \cite{Hamuy_2003} provided a method to calculate $^{56}$Ni mass using the {\em V} band tail photometry and implementing a bolometric correction (BC) factor \citep{hamuy2001PhD} to get the bolometric luminosity. Using this method, the $^{56}$Ni mass was estimated at five nebular epochs (122.28, 127.907, 132.51, 177.51   days). The weighted average of these estimated values results in the $^{56}$Ni mass of 0.0147 $\pm$ 0.0022 M$_\odot$. Another similar approach for estimating the $^{56}$Ni mass is to assume that the SN has the same spectral energy distribution (SED) as the well-studied SN~1987A. In this approach, the $^{56}$Ni mass is obtained by scaling the pseudo-bolometric luminosity of the SN during the nebular phase to that of SN~1987A \citep{Spiro_2014}. Using this method, the $^{56}$Ni mass is calculated at the same five epochs used in the previous method. The weighted average of these estimates yields a $^{56}$Ni mass of 0.013 $\pm$ 0.014 M$_\odot$. Subsequently, the weighted average of the values obtained from the above two methods gives a $^{56}$Ni mass of $0.015 \pm 0.002$ M$_\odot$, similar to SN~2021gmj.

\subsection{Light curve modelling}\label{sec:Lc_modelling}

\subsubsection{Bolometric light curve modelling}\label{sec:Lc_semi_analytical}

\begin{figure}
    \centering
    \includegraphics[width=\columnwidth]{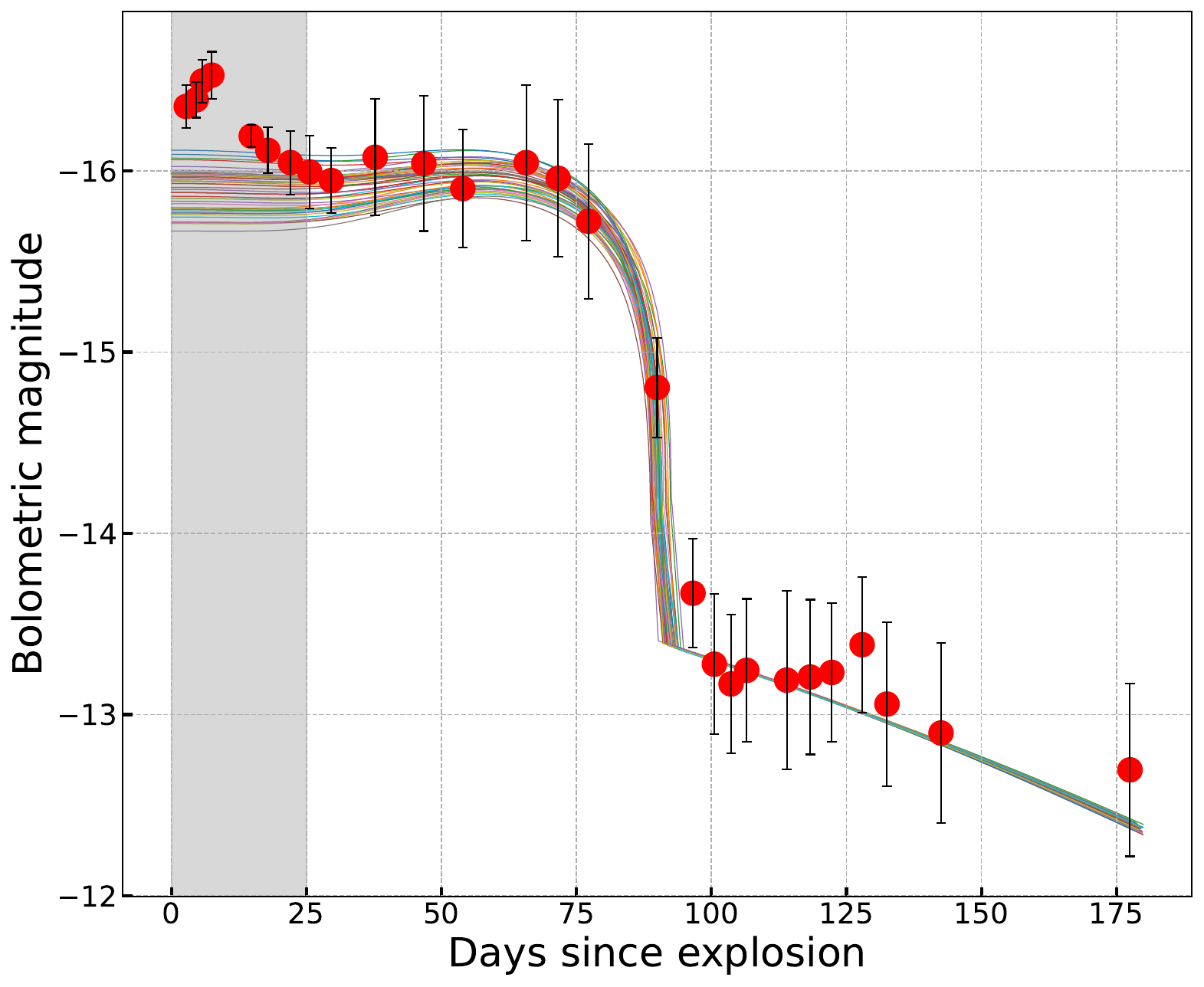}
    \caption{The bolometric light curve of SN~2022xus, along with the 50 best-fit model light curves generated using semi-analytical modelling \citep{Nagy_2014, Jager_2020}, is presented. Data up to 25 days after the explosion (gray shaded region) are not considered, as only the core part is assumed in the model and is accounted for in the modelling.}
    \label{fig:nv}
\end{figure}

\begin{table}
	\begin{center}
	\caption{The best-fit core parameters with 1$\sigma$ uncertainties for the bolometric light curve of SN~2022xus.}
        \label{tab:nv}
	\begin{tabular}{lll} 
		\hline
		Parameter & Best-fit value & Prior range\\
            \hline
            \vspace{0.3cm}
            Initial radius [$R_0$ ($10^{13}$cm)] & ${1.891^{+0.075}_{-0.281}}$ & 1--2 \\
            \vspace{0.2cm}
             Ejecta mass [$M_\mathit{ej}$ (M$_\odot$)] & ${13.067^{+0.017}_{-1.540}}$ & 10--15\\
            \vspace{0.2cm}
            Kinetic energy [$E_k$ (10$^{51}$ ergs)] & ${3.322^{+0.006}_{-0.801}}$ & 2.5--3.5\\
            \vspace{0.2cm}
            Thermal energy [$E_\mathit{th}$ (10$^{51}$ ergs)] & ${0.844^{+0.150}_{-0.051}}$ & 0.01--1\\
            \hline
		
	\end{tabular}
        \end{center}
\end{table}

The progenitor properties can be estimated from the semi-analytical modelling of the bolometric light curve \citep{Nagy_2014}. This model assumes that the spherically symmetric SN ejecta expands homologously, with a constant-density core and an outer envelope of exponentially declining density. Only the core part of this model is considered by \cite{Jager_2020}, who also included the MCMC  routine in the numerical fitting code. The bolometric light curve of SN~2022xus is modelled using \cite{Jager_2020}, considering only the light curve beyond 25 days. The four main progenitor parameters (radius; $R_0$, ejecta mass; $M_{\rm ej}$, kinetic energy; $E_{\rm k}$, and thermal energy; $E_{\rm th}$) are sampled during modelling, while the recombination temperature and Thomson scattering opacity ($\kappa$) are fixed to 7000 K and 0.3 g cm$^{-2}$, respectively. The bolometric and the 50 best-fit model light curves are presented in \autoref{fig:nv}. The model captures the main features of the observed light curve from the onset of the plateau to the nebular phase. The best-fit parameter values and priors used are summarised in \autoref{tab:nv}. The model gives an ejecta mass of $\sim$13 M$_\odot$, which estimates the progenitor mass to be $\sim$15 M$_\odot$, considering the mass of the core compact object as 2 M$_\odot$. This estimation is within the mass range obtained from the nebular spectroscopy (see \autoref{sec:spectra_evo}). The radius of the progenitor from semi-analytical light curve modelling is estimated to be around 258 R$_\odot$, similar to the radius estimated by the shock cooling method in \autoref{sec:shock_cooling}. The estimated total energy (kinetic $+$ thermal) is approximately 4$\times$10$^{51}$ ergs; however, this is higher than the explosion energy found for intermediate-luminous SNe~II ($\sim$ 0.1--0.5 $\times$10$^{51}$ ergs, \citealt{Takats_2009N, Gandhi_2009js}). This energy discrepancy can arise from the simplified assumptions of the model \citep{szalai_2017eaw, raya_2018is, 2018pq_2025}.

\subsubsection{Multi-band light curve modelling}\label{sec:redback}

\begin{figure}
    \centering
    \includegraphics[width=\columnwidth]{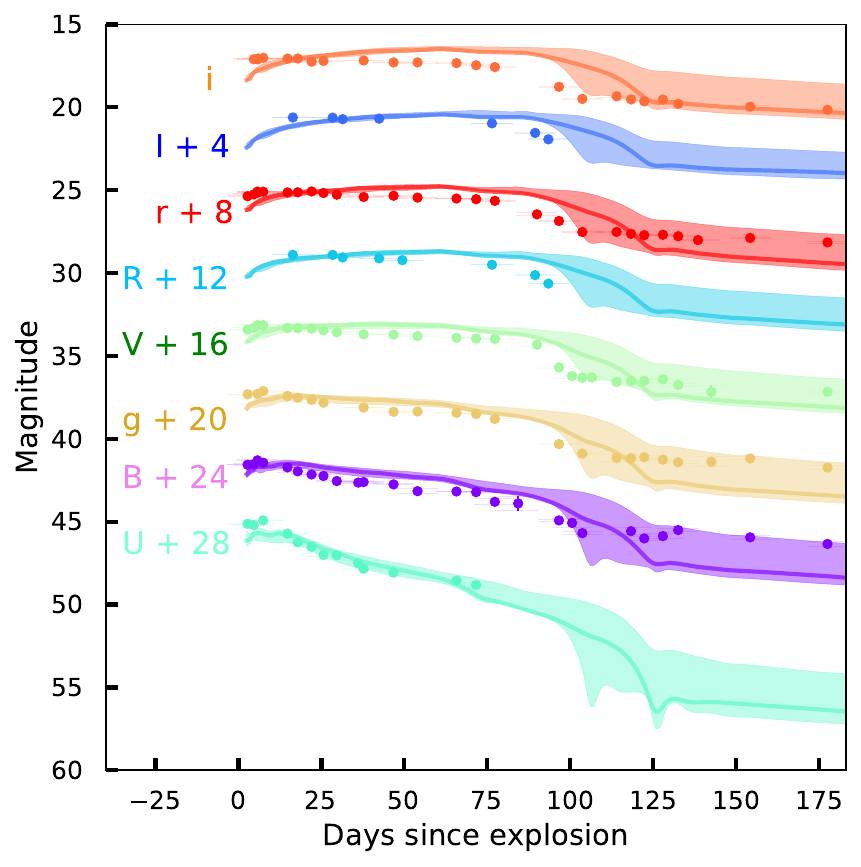}
    \caption{The {\em UBVgRrIi} observed and best fit model light curve. The solid lines show the maximum-likelihood model, while the shaded regions indicate the 99.7\% credible interval ($\approx 3\sigma$) derived from the posterior predictive distribution using \texttt{REDBACK}.}
    \label{fig:redback}
\end{figure}

The multiband light curves of SN~2022xus have been modelled using \texttt{REDBACK} \citep{Sarin_2024, Sarin_2025} to extract the properties of the progenitor. The model has been trained on the large grid of SN~II simulations (228015 simulations) based on RSG progenitor \citep{Sukhbold_2016} provided by \cite{Moriya_2023}. The simulations are performed with the radiation hydrodynamic code \texttt{STELLA} \citep{Blinnikov_1998, Blinnikov_2006}, which provides the photospheric radius and temperature, along with the bolometric light curve and spectral energy distributions (SEDs). The simulation grid from \cite{Moriya_2023} explores a wide parameter space including progenitor zero-age main-sequence mass (ZAMS; 10--18 M$_\odot$), explosion energy (0.5--5 $\times$ 10$^{51}$ ergs), mass of $^{56}$Ni (0.001--0.3 M$_\odot$),  mass-loss rate (10$^{-5}$--10$^{-1}$ M$_\odot$ yr$^{-1}$), and CSM properties such as radial extent (1--10 $\times$ 10$^{14}$ cm) and steepness of density profile (0.5--5). 

In \autoref{fig:redback}, the {\em UBVgri} observed light curves of SN~2022xus are shown with the best fit model light curve with solid lines obtained from \texttt{REDBACK}. The shaded regions show a $\sim 3\sigma$ credible interval derived from the posterior predictive distribution. The input parameters of the model are: ZAMS mass of the progenitor ($M_{\rm ZAMS}$), $^{56}$Ni mass, mass loss rate of the progenitor ($\dot{M}$), steepness of the CSM density profile ($\beta$), radius of the CSM ($R_{\rm CSM}$), and explosion energy ($E_{\rm SN}$). The best-fit values along with the prior ranges are tabulated in \autoref{tab:redback}. During modelling, the explosion epoch is fixed to MJD 59867.61, as previously estimated in \autoref{sec:intro}. The model reproduces all light curves well; however, it slightly overestimates the plateau length and luminosity, which may be due to the fixed value of $^{56}$Ni mixing parameter, which results in $^{56}$Ni being uniformly mixed to the half-mass of the H-rich envelope \citep{Moriya_2023}. The $M_{\rm ZAMS}$ is inferred to be $\sim$ 11.6 M$_\odot$, which is slightly lower than the previously estimated range (12 -- 15 M$_\odot$; see \autoref{sec:spectra_evo} and \autoref{sec:LC_evo}). The $^{56}$Ni mass derived from the surrogate modelling is found to be 0.015 $\pm$ 0.004 M$_\odot$, which is similar to the value estimated from the light-curve analysis presented in \autoref{sec:LC_evo}. A relatively low mass-loss rate of $10^{-4.11}\, \rm M_{\odot}\,\mathrm{yr}^{-1}$ (10$^{-6}$--10$^{-4} \,\rm M_{\odot}\,\mathrm{yr}^{-1}$, for typical RSG star; \citealt{Smith_2014}) is inferred, along with a CSM with $R_{\rm CSM}$ $\sim 4.1 \times 10^{14}$ cm (confined CSM radius for SNe~II is $\sim$ 10$^{14}$--10$^{15}$ cm; \citealt{Morozova_2017}). The steepness of the CSM density profile is estimated as $\sim$ 3.4, indicating a faster fall of the density. These results are consistent with the weak signatures of CSM interaction observed in the early-time spectra of SN~2022xus. The $E_{\rm SN}$ is estimated to be $0.53 \times 10^{51}$~erg, which is significantly lower than the value obtained from semi-analytical modelling in \autoref{sec:LC_evo}. This discrepancy may arise from the simplified assumptions inherent in semi-analytical models, which can lead to a systematic overestimation of the explosion energy.

Overall, the estimated parameters from the multi-band \texttt{REDBACK} agree well with the estimations obtained in previous sections. The surrogate modelling framework allows rapid inference of the progenitor and explosion properties of SN~2022xus, while preserving the predictive accuracy of detailed radiation hydrodynamics simulations at a significantly lower computational cost.

\begin{table}
\centering
\begingroup
\footnotesize
\caption{\texttt{REDBACK} modelling Parameters} 
\label{tab:redback}
\resizebox{\linewidth}{!}{%
\begin{tabular}{ccc}
\hline
Parameter & Prior Range & Best-fit Value \\
\hline
Mass [($M_{\rm ZAMS}$ (M$_\odot$)] & 10 -- 18 & 11.63 $\pm$ 0.57  \\

$^{56}$Ni Mass [$M_{\rm Ni}$(M$_\odot$)] & 0.001 -- 0.3 &  0.015 $\pm$ 0.004  \\

Mass loss rate [$\log_{10} \dot{m}$ (M$_\odot$ yr$^{-1}$)] & $-$5 -- $-$1 & $-$4.11 $\pm$ 0.17 \\

Steepness of the CSM density profile [$\beta$] & 0.5 -- 5.0 & 3.39 $\pm$ 0.46 \\

CSM radius [$R_{\rm CSM}$ ($10^{14}$ cm)] & 1 -- 10 &  4.05 $\pm$ 1.03 \\

Explosion energy [$E_{\rm SN}$ ($10^{51}$ erg)] & 0.5 -- 5.0 &  0.53 $\pm$ 0.01 \\

\hline
\end{tabular}%
}
\endgroup
\end{table}

\section{Assessing the position of the SN in the SNe~II diversity}
\label{sec: diversity}

\begin{figure}
    \centering
    \begin{subfigure}[b]{0.49\textwidth}  
        \centering
        \includegraphics[width=\columnwidth]{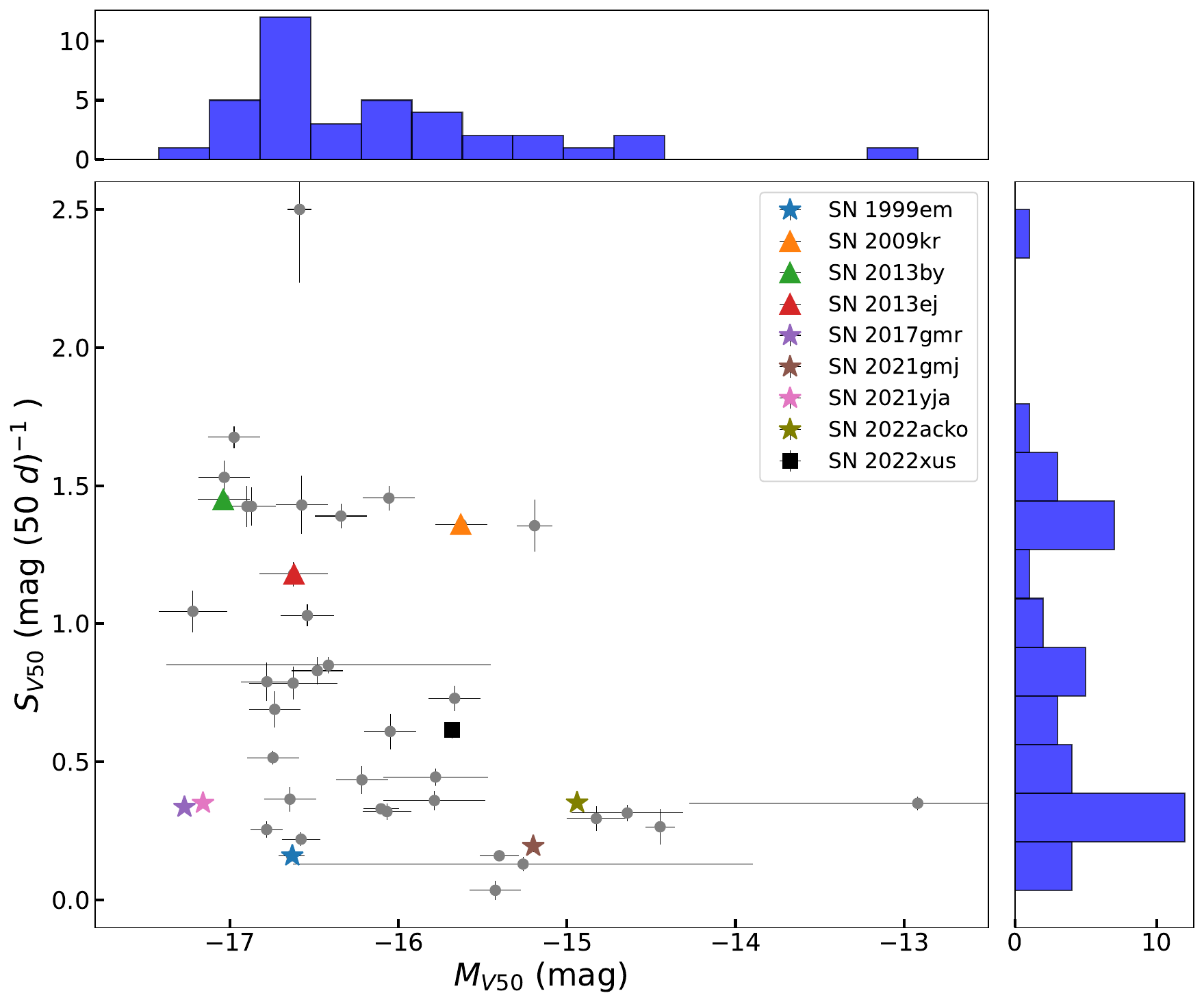}
    \end{subfigure}
    \hfill
    \begin{subfigure}[b]{0.49\textwidth}
        \centering
        \includegraphics[width=\columnwidth]{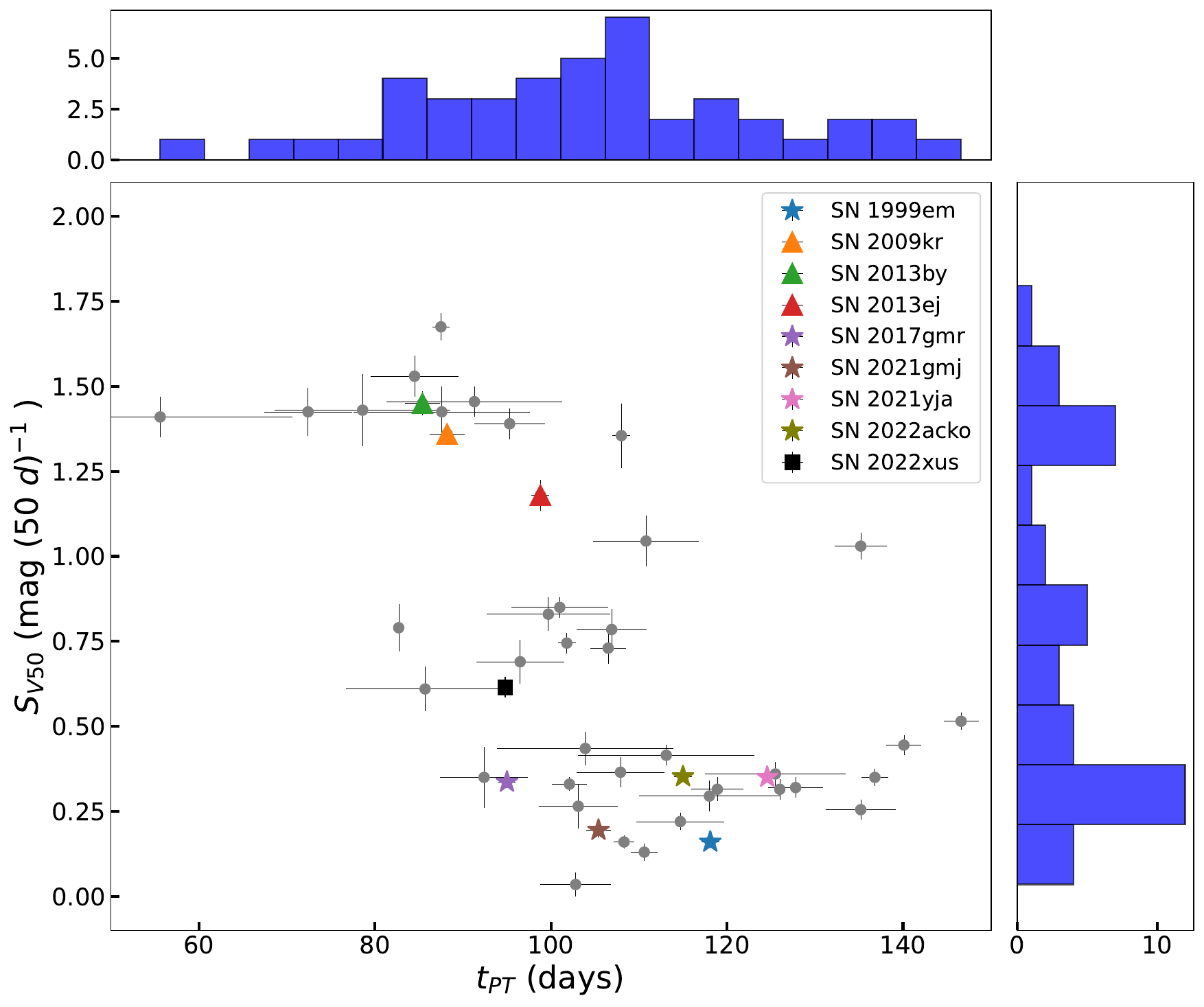}
    \end{subfigure}
    \caption{\textbf{Top Panel:} Correlation plot between $S_{V50}$ and $M_{V50}$, SN~2022xus presents with black square marker, whereas SNe~IIP and IIL are shown in asterisk and triangle marker. An additional SNe~II sample (shown in grey) is taken from \citet{Anderson_2014}. \textbf{Bottom Panel:} The position of SN~2022xus shown in $S_{V50}$--$t_{PT}$ plane.}
    \label{fig:phot-comp}
\end{figure}

\begin{figure*}
    \centering
    \includegraphics[width=\textwidth]{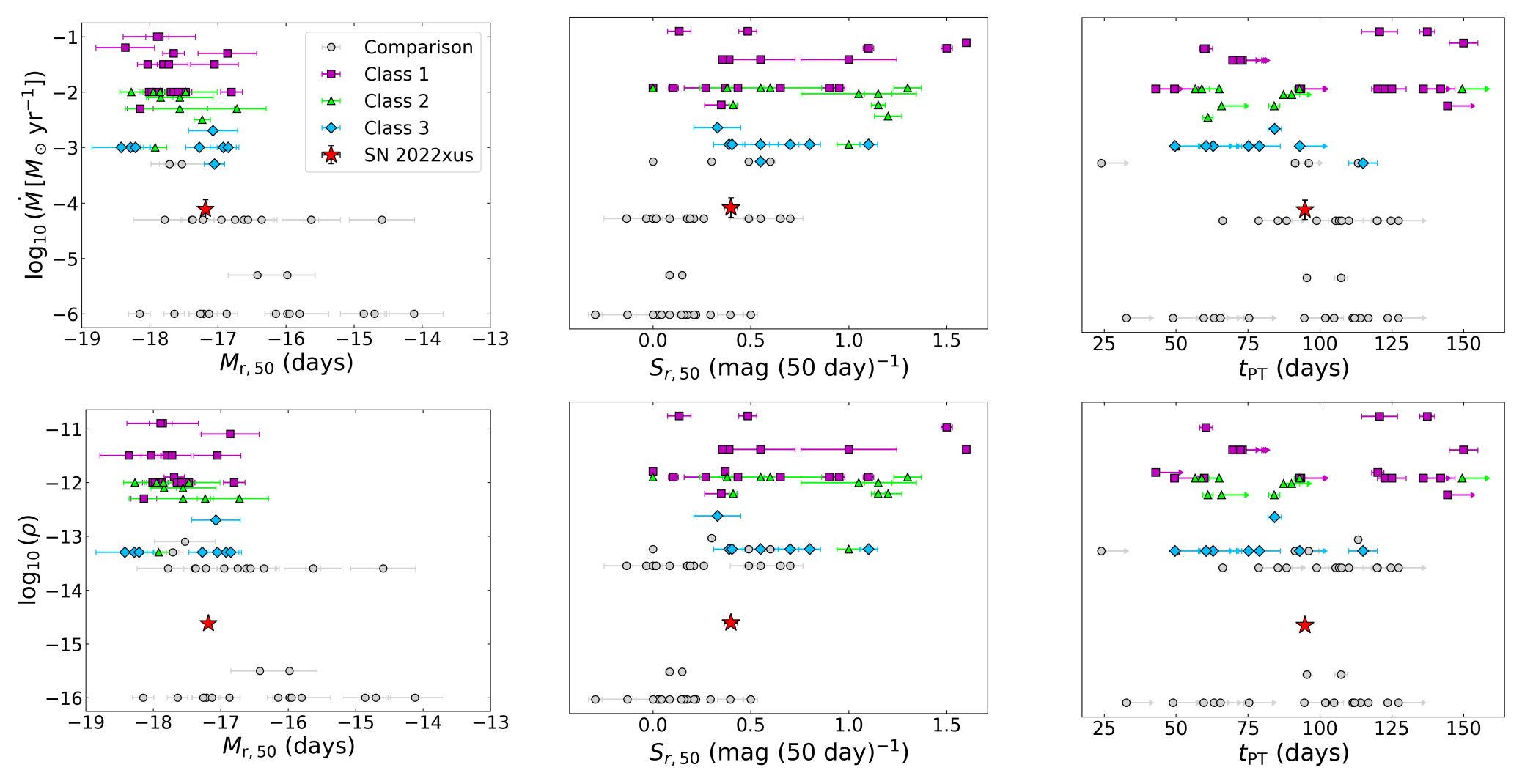}
    \caption{The comparison among the photometric and CSM properties of SN~2022xus (marked with a red star). The additional sample is taken from \citet{Jacobson_2024, Jacobson_2025}, where classes 1, 2, and 3 represent the SNe exhibit flash-ionisation features in decreasing order, along with the comparison SNe with very weak or almost no CSM interaction signature.}
    \label{fig:JG_comp}
\end{figure*}

The detailed photometric and spectroscopic analysis reveals that the SN exhibits characteristics of both SNe~IIP and IIL. Although the SN falls into the intermediate-luminosity category with the peak absolute magnitude in {\em V} band of $-$ 16.32 $\pm$ 0.01 mag, its plateau decline rate ($S_{V}$  $\sim$ 1.2 mag (100 day)$^{-1}$) is significantly steeper than that observed in typical SNe~IIP. The progenitor mass of the SN is estimated in the range of 12--15 M$_\odot$, which is lower than the relatively massive progenitors typically associated with SNe~IIL (e.g., SN~2009kr; \citealt{Elias_2009kr}, SN~2021dgb; \citealt{IIP_IIL_2021dbg}). The SN also shows minimal CSM interaction during the early phase, suggesting low mass loss during the pre-SN phase. However, the early-plateau spectra reveal little to no H$\alpha$ absorption component, implying that the progenitor retained a comparatively thin hydrogen envelope prior to explosion. Hence, to understand the position of SN~2022xus within the diversity of SNe~II, correlations among different SN parameters have been investigated.

The top panel of \autoref{fig:phot-comp} shows a correlation plot between {\em V} band plateau slope per 50 days ($S_{V50}$) and {\em V} band absolute magnitude at 50 day post explosion ($M_{V50}$). In the plot, SN~2022xus is compared with SNe~IIP and IIL of the comparison sample, along with the additional SN sample taken from \citep{Anderson_2014}. In a SNe~II sample study, \citep{Patat_1994} showed that SNe~IIL are brighter than SNe~IIP; however, this distinction has become less evident with increasing sample sizes \citep{Anderson_2014}. In this plot, no significant difference in the absolute magnitude is observed between SNe~IIP and IIL of the comparison sample. SN~2022xus exhibits $M_{V50}$ of $-$15.68 $\pm$ 0.02 mag, similar to SN~IIL 2009kr ($-15.63 \pm 0.15$ mag) and SN~IIP 2021gmj ($-15.20 \pm 0.02$ mag). Although no correlation between $S_{V50}$ and $M_{V50}$ is found, SNe~IIL exhibit higher $S_{V50}$ than SNe~IIP. This could be seen due to low H envelope mass prior to explosion in SNe~IIL, leading to a steeper plateau decline. SNe with $S_{V}$ greater than 1 mag (100 day)$^{-1}$ are typically categorized as SN~IIL \citep{Faran_2014}, and here, the $S_{V}$ of SN~2022xus is 1.23$\pm$0.07 mag (100 day)$^{-1}$, which places it on the SNe~IIL category. In the $S_{V50}$--$t_{PT}$ correlation plot, SNe with longer plateau length show slower plateau decline. For each SNe in the comparison sample, the $t_{PT}$ is measured in a similar manner as described in \autoref{sec:LC_evo}. Most of the SNe~IIP in the comparison sample have a longer plateau duration than SNe~IIL. SN~2022xus exhibits a plateau duration similar to that of the SN~IIL 2013ej and SN~IIP 2017gmr. From these correlation plots, it is observed that the SN shows a steep decline but is not as bright as typical SNe~IIL. Hence, the SN occupies a transitional position, supporting the continuation of characteristics in SNe~IIP and SNe~IIL.

To place SN~2022xus in the context of other SNe~II with early-time flash ionisation signatures, the inferred photospheric and CSM properties of the SN are compared with a sample presented by \citet{Jacobson_2024, Jacobson_2025}. In their study, the sample SNe were classified into three groups (class 1, 2, and 3) based on the duration and spectroscopic evolution of flash-ionisation features. In \autoref{fig:JG_comp}, the absolute {\em r} band magnitude at 50 days ($M_{r,50}$ mag), {\em r} band slope at 50 days ($S_{r,50}$ mag (50 day $^{-1}$)), and the plateau duration ($t_{PT}$ days) are compared with the mass loss rate ($\log_{10}(\dot{M}\,[M_\odot\,{\rm yr^{-1}}])$) and CSM density ($\log_{10}(\rho)$). The density of CSM for SN~2022xus is calculated using the inferred CSM and progenitor properties from the multi-band light curve modelling (see \autoref{sec:redback}) assuming a steady wind, $ \rho = \frac{\dot{M}}{4\pi R_{\rm CSM}^{2}v_{\rm w}}$, where $\dot{M}$ is the mass-loss rate, $R_{\rm CSM}$ is the outer CSM radius, and $v_{\rm w}$ ($10\,\mathrm{km\,s^{-1}}$) is the progenitor wind velocity. Hence, the obtained density,  $\rho = 2.37\times10^{-15}\,\mathrm{g\,cm^{-3}}$, corresponding to $\log_{10}(\rho/\mathrm{g\,cm^{-3}})=-14.62$.

As shown in \autoref{fig:JG_comp}, SN~2022xus exhibits a relatively low mass loss rate of $\log_{10}(\dot{M} \approx -4.11\ M_\odot\,\rm yr^{-1}$), placing it below ($\log_{10}\dot{M} \sim -3.3\ {\rm to} -1.0\ M_\odot\,\rm yr^{-1}$) the flash-ionisation sample of \citet{Jacobson_2024, Jacobson_2025}. The SN lies among the comparison SNe~II sample with a weaker signature of CSM interaction. A similar trend is seen in case of  CSM density, where the estimated $\log_{10}(\rho) \approx -14.6 \,\mathrm{g\,cm^{-3}}$ is significantly lower than the densities found for the flash-ionisation sample ($\log_{10}(\rho) \sim -13\ {\rm to} -11 \,\mathrm{g\,cm^{-3}}$). Although the photometric properties of SN~2022xus are consistent with those of the flash-ionisation sample, the low mass loss rate and CSM density indicate that the SN is embedded in a very low CSM environment.

\begin{table*}
\centering
\renewcommand{\arraystretch}{1.5} % Adjust the spacing
    \caption{Properties of the SN~2022xus and the comparison sample SNe.}
    \label{tab:comp_SN}
    \resizebox{\linewidth}{!}{%
    \begin{tabular}{ccccccccccc}
    \hline 
    Supernova & \makecell{Parent \\ Galaxy} & \makecell{Distance \\ (Mpc)} & \makecell{$E(B-V)_{tot}$ \\ (mag)} & \makecell{$M_{V50}$ \\ (mag)} & \makecell{$t_\mathit{tp}$ \\ (days)} & \makecell{$E$($10^{51}$) \\ (ergs)} & \makecell{$R$ \\ ($R_\odot$)} & \makecell{$M_\mathit{ej}$ \\ ($M_\odot$)} & \makecell{$^{56}$Ni \\ ($M_\odot$)} & Ref. \\
    \hline

    1990K & NGC 150 & 21.61$\pm$1.52 & 0.5$\pm$0.1 & $-15.00\pm0.04$ & -- & -- & $\sim$431 & -- & $\sim$0.1 & 1 \\
    
    1999em & NGC 1637 & 11.7$\pm$0.1 & 0.1 & $-16.63\pm0.07$ & 118.1$\pm$1.0 & $0.5-1$ & $120-150$ & $10-11$ & $0.042^{+0.027}_{-0.019}$ & 2, 3, 4 \\

    2009kr & NGC 1832 & 26.20$\pm$1.84 & 0.08$\pm$0.01 & $-15.63\pm0.15$ & 88.2$\pm$2.0 & -- & -- & 18--24 ($M_{ZAMS}$) & 0.009$\pm$0.004 & 5 \\

    2013by &  ESO 138 G10 & 14.50$\pm$1.03 & 0.195 & $-17.04\pm0.15$ & 85.4$\pm$2.0 & $\sim$1.4 & $\sim$2300 & 14.5 ($M_{ZAMS}$) & 0.029$\pm$0.005 & 6,7 \\

    2013ej & NGC 628 & 9.93$\pm$0.71 & 0.060$\pm$0.001 & $-16.62\pm0.20$ & 98.8$\pm$1.0 & $\sim$2.3 & $\sim$450 & $\sim$14 ($M_{ZAMS}$) & 0.019$\pm$0.002 & 8\\

    2017gmr & NGC 988 & 19.92$\pm$1.40 & 0.30 & $-17.27\pm0.03$ & 95.1$\pm$1.0 & 10.2 & 525 & 22 & 0.130$\pm$0.026 & 9,10\\

    2021gmj & NGC 3310 & $19.12\pm1.36$ & 0.049$\pm$0.011 & $-15.20\pm0.02$ & 105.4$\pm$1.4  & 0.294 & -- & 10 & 0.014$\pm$0.001 & 11,12 \\  

    2021yja &  NGC 1325 &  $20.89\pm1.47$ & 0.104 & $-17.16\pm0.03$ & 124.6$\pm$0.5  & 1.53 & 631 & 15 ($M_{ZAMS}$) & 0.175--0.2 & 13, 14\\

    2022acko & NGC 1300 & 20.72$\pm$1.45 & 0.04$\pm$0.01 & $-14.94\pm0.01$  & 115.0$\pm$1.0 & $\sim$0.26--0.5 & $\sim$400--500 & $\sim$5--7 &  0.017$^{+0.009}_{-0.007}$ & 15, 16   \\

    \bfseries 2022xus & \bfseries LEDA 136560 & \bfseries 37.43$\pm$2.63 & \bfseries 0.193 & $\mathbf{-15.68\pm0.02}$ & \bfseries 94.8 $\pm$0.4 & \bfseries $\sim$ 4 & \bfseries $\sim$ 270 & \bfseries $\sim$13  & \bfseries  $\mathbf{0.015\pm0.002}$ & \bfseries This work \\  
    \hline
    \end{tabular}
    }
    References: (1) \cite{Cappellaro_1990K}, (2) \cite{Hamuy_2001}, (3) \cite{Leonard_2002}, (4) \cite{Elmhamdi_2003}, (5) \cite{Elias_2009kr}, (6) \cite{valenti_2013by}, (7) \cite{Morozova_2013by_2013ej}, (8) \cite{bose_2013ej}, (9) \cite{Andrews_2019}, (10) \cite{Utrobin_2017gmr}, (11) \cite{Murai_2021gmj}, (12) \cite{Meza_2021gmj_2024}, (13) \cite{Kozyreva_2021yja_2022}, (14) \cite{Hosseinzadeh_2021yja_2022}, (15) \cite{Azalee_2022acko}, (16) \cite{Lin_2022acko}

\end{table*}

\begin{figure}
    \centering
    \includegraphics[width=\columnwidth]{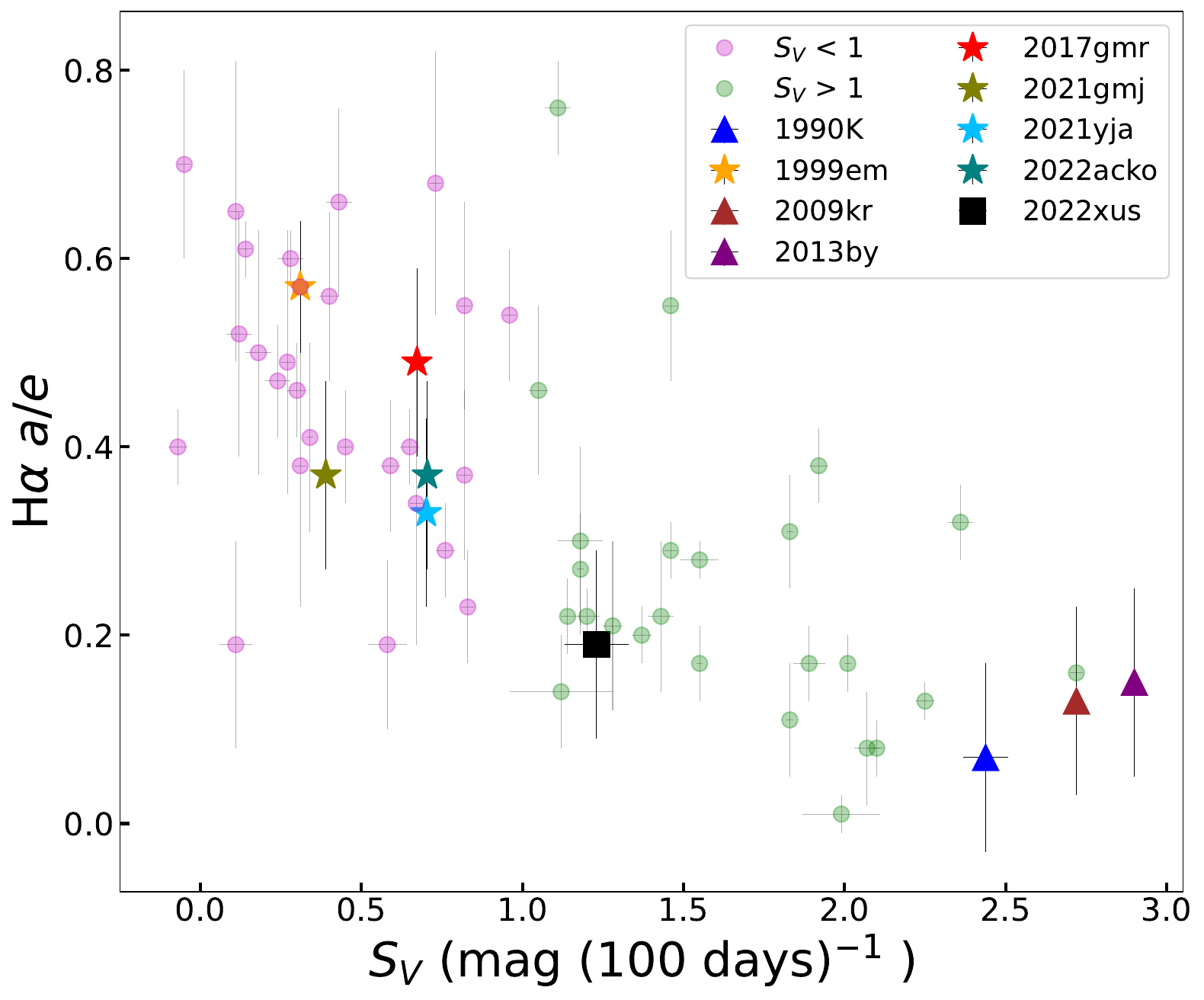}
    \caption{The correlation between $S_V$ and H$\alpha$ $a/e$ ratio is shown. Circular markers denote the SNe from \citet{Gutierrez_2014}, with magenta and green colour indicating $S_V$ values greater or less than 1. The SNe~IIP and IIL of the comparison sample are plotted as asterisks and triangles, respectively, while SN~2022xus is represented by a black square.}
    \label{fig:EW}
\end{figure}

In spectral comparison, SN~2022xus does not show a prominent H$\alpha$ absorption dip similar to typical SNe~IIL, in the early plateau phase (see \autoref{fig:spec_ev}, right panel). To compare the amount of H in the SN with other SNe, the H$\alpha$ absorption-to-emission ($a/e$) ratio during the mid plateau phase is plotted against $S_V$ in \autoref{fig:EW}. The plot includes SN~2022xus, SNe~IIP, and SNe~IIL from the comparison SNe, and an additional sample from \cite{Gutierrez_2014}, in which the green and magenta colours represent the SNe with $S_V$ value greater or less than 1, respectively. From the spectroscopic study on 52 SNe, \cite{Gutierrez_2014} found an anti-correlation between H$\alpha$ $a/e$ ratio and $S_V$, indicating faster declining SNe tend to have progenitors with a low mass H envelope prior to the explosion. The typical SNe~IIP and IIL in the comparison sample also show the same trend. The H$\alpha$ $a/e$ ratio of SN~2022xus lies between that of the typical SNe~IIP and IIL of the comparison sample. The plot does not clearly distinguish between SNe~IIP and IIL in the larger sample; rather, it indicates a continuum between the two populations, reflecting variations in the H envelope mass of the progenitor, which in turn influences the plateau decline rate. The position of SN~2022xus is consistent with this trend.

\begin{figure}
    \centering
    \begin{subfigure}[b]{0.49\textwidth}  
        \centering
        \includegraphics[width=\columnwidth]{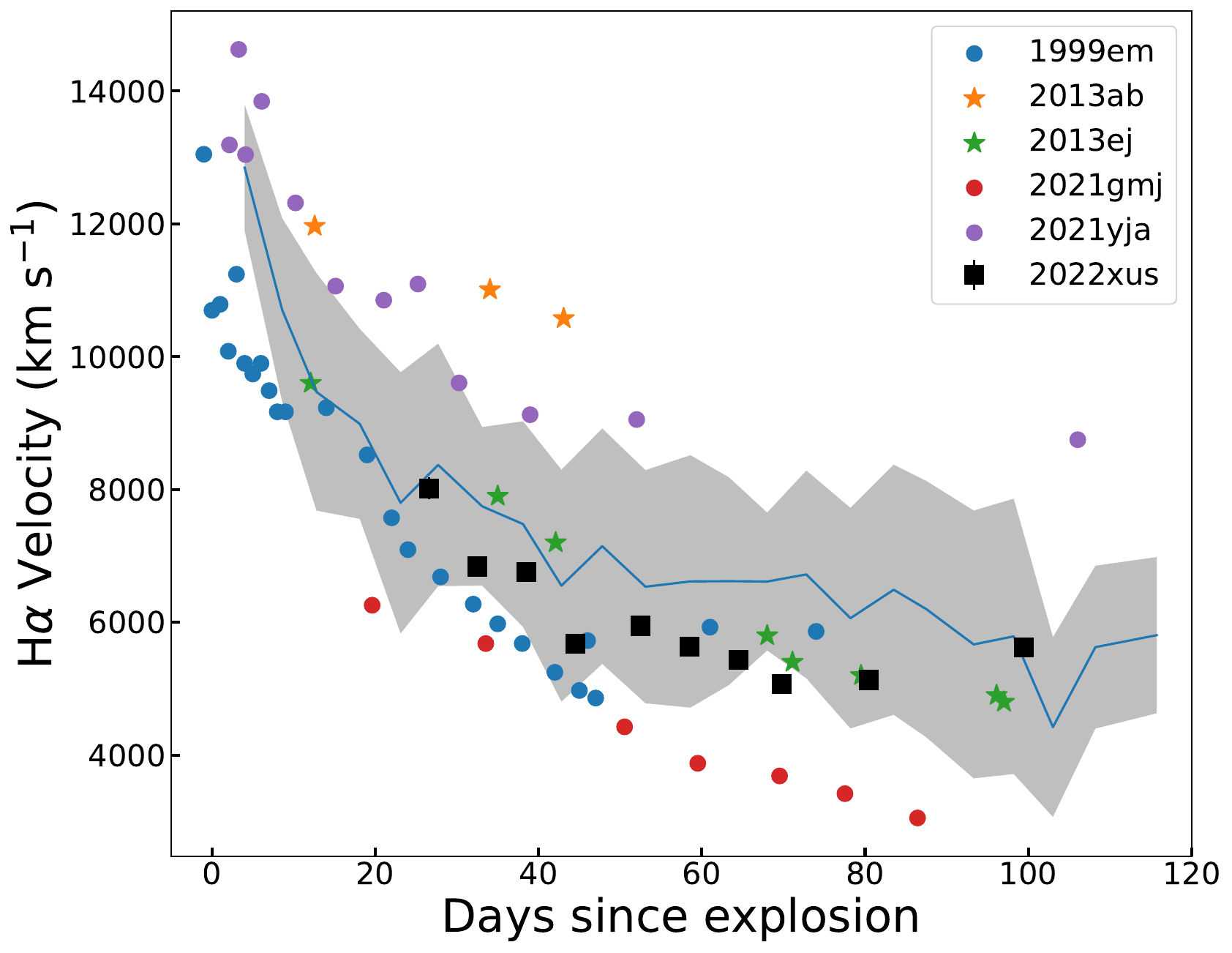}
    \end{subfigure}
    \hfill
    \begin{subfigure}[b]{0.47\textwidth}
        \centering
        \includegraphics[width=\columnwidth]{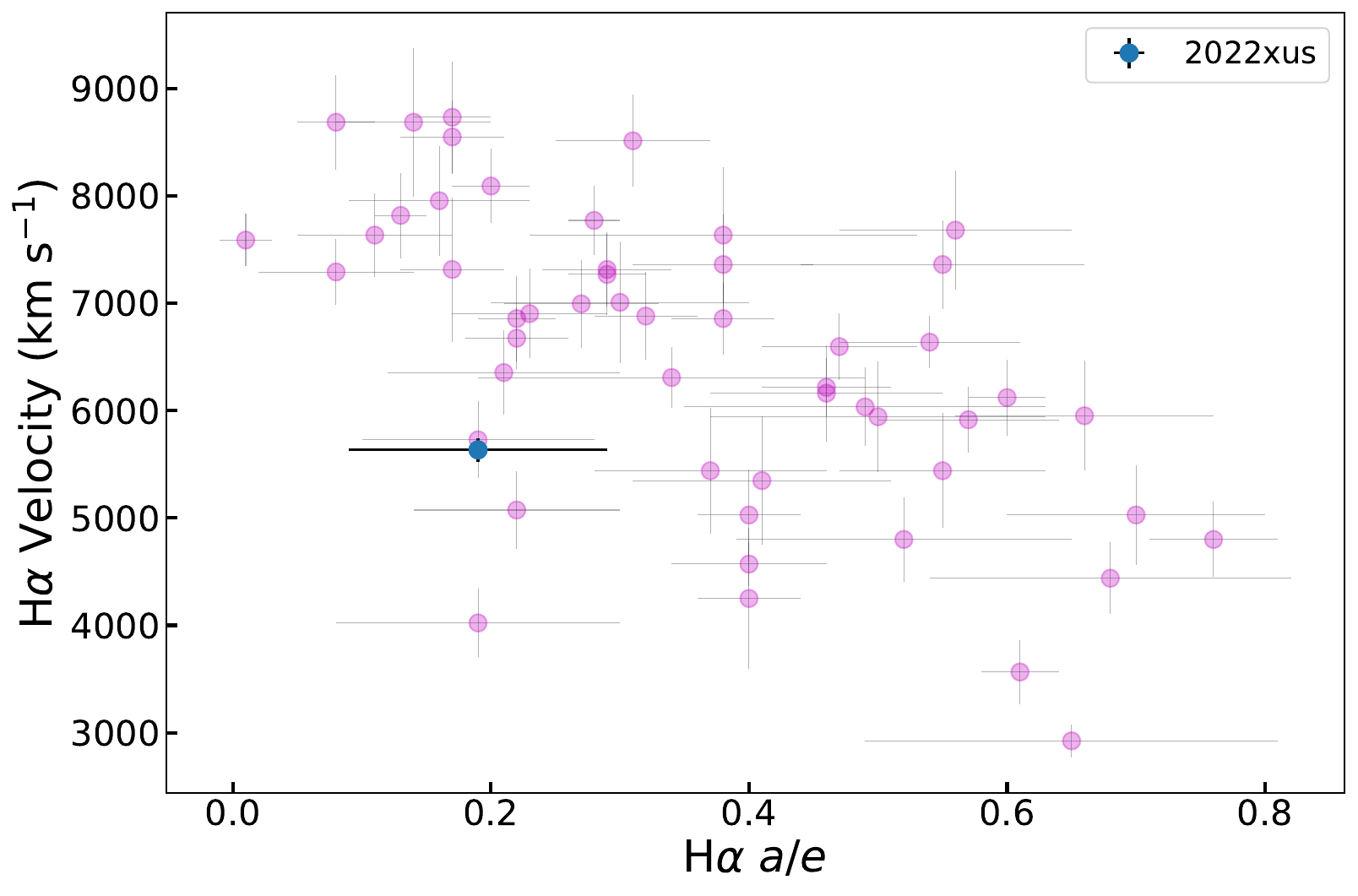}
    \end{subfigure}
    \caption{\textbf{Top Panel:} The velocity evolution of SN~2022xus (black square) is plotted together with five SNe from the comparison sample, where SNe IIP and IIL are shown as circles and asterisks, respectively. The blue line shows the mean velocity from the extended sample of \citet{Gutierrez_2014}, with its uncertainty shaded in gray. \textbf{Bottom Panel:} The location of SN~2022xus is displayed in the H$\alpha$ velocity--$a/e$ ratio plane, alongside the magenta points representing the sample from \citet{Gutierrez_2014}.}
    \label{fig:vel-comp}
\end{figure}

In \autoref{fig:vel-comp} (top panel), the H$\alpha$ velocity evolution of SN~2022xus is plotted along with the comparison SNe, as well as the mean velocity and the corresponding uncertainty of the SNe~II sample study done by \cite{Gutierrez_2014}. No clear distinction is found between the SNe~IIP and IIL. SNe~2013ab (IIL) and 2021yja (IIP) have a higher velocity, while SN~2022xus shows a velocity lower than the mean, similar to SN~2013ej. The prototypical SN~IIP SNe~1999em and 2021gmj have lower velocities than the SN. \cite{Gutierrez_2014} found an anti-correlation between the H$\alpha$ velocity with the $a/e$ ratio during the photospheric phase, which is shown in \autoref{fig:vel-comp} (bottom panel). This suggests that smaller H$\alpha$ $a/e$ ratio SNe have higher velocities and decline more rapidly. These SNe are historically classified as SNe~IIL; however, the continuation in the properties of SNe~IIP and IIL is also evident here. SN~2022xus shows a velocity that is close to the average of the distribution, residing toward the lower boundary of the $a/e$ ratio.

From the present analysis, we conclude that SN~2022xus exhibits an average brightness and a moderate expansion velocity, as seen in many SNe~IIP; however, the steeper plateau decline and lower H$\alpha$ $a/e$ ratio indicate characteristics more similar to those of SNe~IIL. Therefore, the SN behaves as a transitional candidate between these two categories, further supporting the continuity between the two populations. 

\section{Summary and conclusions}
\label{sec:summary}
In this work, we present the multi-band photometric and spectroscopic analysis of SN~2022xus. The SN was discovered by ATLAS $\sim$1 day after the explosion and was followed up until the late nebular phase ($\sim$422 days). The important findings from our analysis are summarised below.

\begin{enumerate} 
    \item SN~2022xus is categorised as SNe~II with a prominent H$\alpha$ P-Cygni profile visible during the spectroscopic evolution. The absence of high-ionisation lines in the earliest spectrum, along with the appearance of the `ledge' feature, supports low-level CSM interaction during the early evolution of the SN.

    \item As the SN evolves, it displays features of both SNe~IIP and IIL in its spectroscopic evolution. In the early-plateau phase, it exhibits broad Balmer emission lines with nearly zero absorption, similar to SNe~IIL. The lower H$\alpha$ $a/e$ ratio during the photospheric phase further suggests the presence of a low-mass H envelope, unlike a typical SN~IIP. From the nebular spectroscopy, the mass of the progenitor is constrained within the range of 12--15 M$_\odot$. 

    \item The SN reached a peak magnitude of $-16.32 \pm 0.01$ at $\sim$7 days since explosion in {\em V} band, corresponding to an intermediate-luminosity event. The plateau phase lasts for $94.80 \pm 0.44$ days with a decline rate of $1.23 \pm 0.07$ mag (100 d)$^{-1}$. The estimated $^{56}$Ni mass obtained from the tail luminosity is 0.015$\pm$0.002 M$_\odot$.

    \item The semi-analytical modelling of the bolometric light curve estimates the progenitor mass to be $\sim$15 M$_\odot$, which is in good agreement with the result from nebular spectroscopy. The multi-band modelling of the SN has been performed using \texttt{REDBACK}, trained on a large grid of simulations computed with \texttt{STELLA}. The inferred progenitor mass is approximately 11.63 M$_\odot$, with a $^{56}$Ni mass of 0.015 M$_\odot$; both of them are consistent with the estimates obtained in earlier analyses. The low mass-loss rate and compact CSM are indicative of an RSG progenitor.

    \item The correlation studies among the different parameters derived from light curves and spectra analysis suggest there is no strict boundary between SNe~IIP and IIL. SN~2022xus lies between these two populations, supporting their continuity \citep{Anderson_2014, Gutierrez_2014, Valenti_2016}.
\end{enumerate}

Based on the overall analysis of SN~2022xus, we conclude that it is a transitional event. The well-cadenced photometric and spectroscopic observations enable us to robustly constrain the properties across the different phases of SN evolution. The SN exhibits characteristics of both subclasses, making it a unique event within the SNe~IIP and IIL populations. This work also highlights the need for high-cadence observations in SN studies, as they are crucial for capturing rapid evolution, especially in the early phases. Increasing the sample of such well-observed events will be critical for studying the continuum of light-curve morphologies, as well as the significance of the H envelope, the initial mass, and mass-loss history of the progenitor in governing SN evolution.

\section*{Acknowledgements}
We thank the referee for providing constructive comments on the manuscript, which improved the content and clarity of the presentation. This work uses data from the Las Cumbres Observatory (LCO) Global Telescope Network. The LCO group is supported by U.S. National Science Foundation (NSF) grants AST-2308113 and AST-1911151. This paper includes data obtained with the ARIES 130-cm Devasthal Fast Optical Telescope (DFOT). M.D. acknowledges the Innovation in Science Pursuit for Inspired Research (INSPIRE) fellowship award (DST/INSPIRE Fellowship/2020/IF200251) for this work. K.M. and N. D. acknowledge the support from the BRICS grant DST/ICD/BRICS/Call-5/CoNMuTraMO/2023 (G) funded by the Department of Science and Technology (DST), India. K.A.B. is supported by an LSST-DA Catalyst Fellowship. S.V. and the UC Davis time-domain research team acknowledge support by NSF grants AST-2407565. In addition, one spectrum presented herein was obtained at Keck Observatory on Maunakea, which is a private 501(c)3 nonprofit organization operated as a scientific partnership among the California Institute of Technology, the University of California, and the National Aeronautics and Space Administration. The Observatory was made possible by the generous financial support of the W. M. Keck Foundation. 
%%%%%%%%%%%%%%%%%%%%%%%%%%%%%%%%%%%%%%%%%%%%%%%%%%
\section*{Data Availability}
 The data presented in this paper are available upon request. The spectra will be publicly accessible via Zenodo.



\bibliographystyle{mnras}
\bibliography{2022xus_reference} % if your bibtex file is called example.bib

%\begin{thebibliography}{99}
%\end{thebibliography}

%%%%%%%%%%%%%%%%%%%%%%%%%%%%%%%%%%%%%%%%%%%%%%%%%%

%%%%%%%%%%%%%%%%% APPENDICES %%%%%%%%%%%%%%%%%%%%%

%\appendix

%\section{Photometric Data}

%%%%%%%%%%%%%%%%%%%%%%%%%%%%%%%%%%%%%%%%%%%%%%%%%%

% Don't change these lines
\bsp	% typesetting comment
\label{lastpage}
\end{document}